\documentclass[preprint,eqsecnum,prb]{revtex4}
\usepackage{epsfig,amsmath,bbm,graphicx}

\newcommand{\ee}{\end{equation}}
\newcommand{\be}{\begin{equation}}
\newcommand{\bea}{\begin{eqnarray}}
\newcommand{\eea}{\end{eqnarray}}
\newcommand{\bml}{\begin{subequations}} 
\newcommand{\eml}{\end{subequations}}

\begin{document}

\title{Numerically exact, time-dependent treatment of vibrationally coupled 
electron transport in single-molecule junctions}

\author{Haobin Wang}
\affiliation{Department of Chemistry and Biochemistry, MSC 3C, New Mexico State
University, Las Cruces, NM 88003}
\author{Ivan Pshenichnyuk, Rainer H\"artle, Michael Thoss}
\affiliation{Institut  f\"ur Theoretische Physik und 
Interdisziplin\"ares Zentrum f\"ur Molekulare Materialien,
  Friedrich-Alexander-Universit\"at Erlangen-N\"urnberg,
  Staudtstr.\ 7/B2, D-91058, Germany}


\begin{abstract}
\baselineskip6mm

The multilayer multiconfiguration time-dependent Hartree (ML-MCTDH) theory within second 
quantization representation of the Fock space, a novel numerically exact methodology to treat
many-body quantum dynamics for systems containing identical particles, is
applied to study the effect of vibrational motion on electron transport in
a generic model for single-molecule junctions.  
The results demonstrate the importance of electronic-vibrational coupling for
the transport characteristics.
For situations where the energy of the bridge state is located close to the
Fermi energy, the simulations show the time-dependent formation of a polaron
state that results in a pronounced suppression of the current corresponding to
the phenomenon of phonon blockade. 
We show that this phenomenon cannot be explained solely by the polaron shift
of the energy but requires methods that incorporate the dynamical effect of
the vibrations on the transport. The accurate results obtained with the
ML-MCTDH in this parameter regime are compared to results of nonequilibrium 
Green's function (NEGF) theory.

\end{abstract}
\maketitle

\section{Introduction}

Charge transport in single-molecule junctions, i.e.\ single-molecules that are
bound to metal or semiconductor electrodes, has been of great interest
recently.\cite{ree97:252,joa00:541,Nitzan01,nit03:1384,Cuniberti05,Selzer06,Venkataraman06,Chen07,Galperin08b,Cuevas10}
Employing different experimental techniques, including electromigration or
mechanically controllable break junctions or  scanning tunneling
microscopy,\cite{ree97:252,par00:57,cui01:571,par02:722,smi02:906,rei02:176804,zhi02:226801,xu03:1221,qiu04:206102,liu04:11371,elb05:8815,Elbing05,Ogawa07,Schulze08,Pump08,Leon08,Osorio10,Tao10,Martin10}
the conductance properties of nanoscale molecular junctions have been
investigated. The observed current-voltage characteristics typically exhibit 
 a nonlinear behavior with resonance structures at larger bias voltages associated with the discrete energy 
levels of the molecular bridge.
The experiments have also revealed a wealth of interesting transport phenomena
including  Coulomb blockade,\cite{par02:722} the Kondo effect,\cite{lia02:725} negative differential 
resistance,\cite{che99:1550,Gaudioso00,Osorio10} switching and hysteresis.\cite{blu05:167,Riel06,Choi06}
Furthermore the possibility to obtain transport characteristics that resemble
those of a diode\cite{elb05:8815} or a transistor\cite{par00:57} has been demonstrated. 
 These findings have stimulated great interest in
the basic mechanisms which govern quantum transport 
at the molecular scale.

An interesting aspect that distinguishes single-molecule junctions from mesoscopic devices is the influence 
of nuclear motion on electron transport. Because of the small size  of
molecules,  the charging of the molecular bridge 
is often accompanied
by significant changes of the nuclear geometry that result in
strong coupling between electronic and vibrational degrees of freedom. 
This coupling may give rise to substantial current-induced vibrational
excitation and thus may cause heating and possible breakage of the molecular junction.  
The signature of nuclear motion has been observed in conduction measurements
of a variety of molecular junctions,\cite{sti98:1732,par00:57,smi02:906,zhi02:226801,qiu04:206102,liu04:11371,Natelson04,kus04:639,pas05:203,Pasupathy05,Sapmaz06,Thijssen06,Parks07,Boehler07,Leon08,Huettel09,Tao10,Ballmann10,Osorio10,Secker11}
e.g., H$_2$ between platinum electrodes,\cite{smi02:906} C$_{60}$ molecules between gold 
electrodes,\cite{par00:57} 
and copper phthalocyanine\cite{liu04:11371} on aluminum oxide film. 
Vibrational signatures of molecular bridges have also been observed in inelastic electron 
tunneling spectroscopy.\cite{sti98:1732,kus04:639,pas05:203}
New experimental techniques\cite{Huang06,Natelson08,Ioffe08} based, e.g., on
Raman spectroscopy, allow the  characterization of the
nonequilibrium state of the vibrational degrees of freedom in a molecular junction.

The experimental progress has stimulated much interest
in the theoretical modeling and simulation of vibrationally
coupled electron transport in molecular junctions.
To this end, a variety of theoretical approaches have been developed and
employed, including scattering
theory,\cite{Bonca95,Ness01,Cizek04,Cizek05,Toroker07,Benesch08,Zimbovskaya09,Seidemann10} 
nonequilibrium Green's function
approaches,\cite{Flensberg03,Mitra04,Galperin06,Ryndyk06,Frederiksen07,Tahir08,Haertle08,Stafford09,Haertle09} 
and master equation methods.\cite{May02,Mitra04,Lehmann04,Pedersen05,Harbola06,Zazunov06,Siddiqui07,Timm08,May08,May08b,Leijnse09,Esposito09,Haertle11} Although much physical insight has been
obtained by the application of these methods, all these approaches involve
significant approximations. For example,  NEGF methods and
master equation approaches are usually based on (self-consistent) perturbation theory and/or employ
factorization schemes. Scattering
theory approaches to vibrationally coupled electron transport, on the other
hand, 
neglect vibrational nonequilibrium effects and are limited to the
treatment of a small number of vibrational degrees of freedom. 
Furthermore, a systematic improvement of these approaches to yield
numerically exact result, though
formally possible by, e.g., including higher orders in the perturbation
expansion, is practically very challenging.
These shortcomings have motivated us to develop a
systematic, numerically exact methodology
to study quantum dynamics and quantum transport including 
many-body effects, in particular, correlated electronic-nuclear dynamics  ---
the multilayer multiconfiguration time-dependent Hartree (ML-MCTDH) theory 
in second quantization 
representation (SQR).\cite{wan09:024114}   
Other efforts along the same direction include the numerical path integral
approach,\cite{muh08:176403,wei08:195316,Segal10} real-time quantum Monte Carlo 
simulations,\cite{Werner09,Schiro09} the numerical renormalization 
group approach,\cite{and08:066804} and the time-dependent density matrix
renormalization group.\cite{HeidrichMeisner09} For a comparison and an overview of various different
methods in the related problem of nonequilibrium transport with
electron-electron interaction see Ref.\ \onlinecite{Eckel10}.

In this paper, we report results of accurate quantum simulations employing the
ML-MCTDH-SQR theory
for a generic model of vibrationally
coupled
electron transport through molecular junctions.  The paper is organized as
follows. Section~\ref{modeltight} outlines the
physical model and the observables of interest.  The ML-MCTDH-SQR theory is
described in Section~\ref{mlsqr}.  Section~\ref{results} presents numerical
results for vibrationally coupled electron transport in different parameter
regimes as well as an analysis of the transport mechanism. Moreover, the
validity of NEGF theory in the regime of phonon blockade is discussed.
Finally, Section~\ref{conclusions} concludes.

\section{Model and Observables of Interest}\label{modeltight}

To study vibrationally coupled electron transport we consider a simple generic model
for a single-molecule junction.  It comprises one discrete electronic state at 
the molecular junction, two electronic continua describing the left and the right metal leads,
respectively, and a distribution of harmonic oscillators that models the
vibrational modes of the molecular bridge.  
The Hamiltonian  reads
\begin{subequations}\label{Htot}
\begin{equation}
	\hat H = \hat H_{\rm el} + \hat H_{\rm nuc} + \hat H_{\rm el-nuc},
\end{equation}
where $\hat H_{\rm el}$, $\hat H_{\rm nuc}$, and $\hat H_{\rm el-nuc}$ 
describe the electronic degrees of freedom, the nuclear
vibrations, and their coupling terms, respectively
\begin{eqnarray}
	\hat H_{\rm el} &=& E_d d^+ d + \sum_{k_L} E_{k_L} c_{k_L}^+ c_{k_L}
	+ \sum_{k_R} E_{k_R} c_{k_R}^+ c_{k_R} \\
  &&	+ \sum_{k_L} V_{dk_L} ( d^+ c_{k_L} + c_{k_L}^+ d )
	+ \sum_{k_R} V_{dk_R} ( d^+ c_{k_R} + c_{k_R}^+ d ), \nonumber \\
	\hat H_{\rm nuc} &=& \frac{1}{2} \sum_j ( P_j^2 + \omega_j^2 Q_j^2 ), \\
	\hat H_{\rm el-nuc} &=& d^+ d \sum_j c_j Q_j.
\end{eqnarray}
\end{subequations}
Thereby,
$d^+/ d$, $c_{k_L}^+/ c_{k_L}$, $c_{k_R}^+/ c_{k_R}$ are the
fermionic creation/annihilation operators for the electronic states on the molecular bridge, the left
and the right leads, respectively.  The corresponding electronic energies $E_{k_L}$, $E_{k_R}$ 
and the molecule-lead coupling strengths $V_{dk_L}$, $V_{dk_R}$, are defined through the energy-dependent
level width functions
\begin{equation}
	\Gamma_L (E) = 2\pi \sum_{k_l} |V_{dk_L}|^2 \delta(E-E_{k_L}), \hspace{1cm}
	\Gamma_R (E) = 2\pi \sum_{k_r} |V_{dk_R}|^2 \delta(E-E_{k_R}).
\end{equation}
In principle, the parameters of the model can be obtained for a specific
molecular junction employing first-principles electronic structure
calculations.\cite{Benesch09} In this paper, which focuses on the transport
methodology, however, we will use a generic parametrization.
Employing a tight-binding model, the function $\Gamma (E)$ is given as
\begin{subequations}
\begin{equation}
        \Gamma (E) = \left\{ \begin{array}{ll} \frac{\alpha_e^2}{\beta_e^2} \sqrt{4\beta_e^2-E^2} 
		\hspace{1cm} & |E| \leq 2 |\beta_e| \\
                0 \hspace{1cm} &  |E| > 2 |\beta_e| \end{array}  \right.,
\end{equation}
\begin{equation}
	\Gamma_L (E) = \Gamma (E-\mu_L), \hspace{1cm}  \Gamma_R (E) = \Gamma (E-\mu_R),
\end{equation}
\end{subequations}
where $\beta_e$ and $\alpha_e$ are nearest-neighbor couplings between two lead sites 
and between the lead and the bridge state, respectively.  I.e., the width functions for 
the left and the right leads are obtained by shifting $\Gamma(E)$ relative to the chemical potentials 
of the corresponding leads.  We consider a simple case of two identical leads, in which the chemical 
potentials are given by 
\begin{equation}
	\mu_{L/R} = E_f \pm V/2,
\end{equation}
where $V$ is the bias voltage and $E_f$ the Fermi energy of the leads.  Since only the difference $E_d - E_f$ 
is physically relevant, we set $E_f = 0$ in this paper.

The frequencies $\omega_j$ and electronic-nuclear coupling
constants $c_j$ of the vibrational modes of the molecular junctions are modeled by a spectral density 
function\cite{leg87:1,Weiss93}
\begin{equation}
\label{discrete}
        J(\omega) = \frac{\pi} {2} \sum_{j} \frac{c_{j}^{2}} {\omega_{j}}
        \delta(\omega - \omega_{j}).
\end{equation}
In this paper, the spectral density is chosen in  Ohmic form with an exponential cutoff
\begin{equation}
\label{ohmic}
        J_{\rm O}(\omega)  = \frac{\pi}{2} \alpha \omega e^{-\omega/\omega_c},
\end{equation}
where $\alpha$ is the dimensionless Kondo parameter.  

Both the electronic and the vibrational continua can
be discretized using an appropriate scheme.\cite{wan03:1289}  Within a given time scale the numbers 
of electronic states and bath modes are systematically increased to reach converged
results for the quantum dynamics in the condensed phase.  In this paper, we employ 64-128 states
for each electronic lead, implying 32-64 electrons per lead, and a bath with 100-400 modes.

The observable of interest in transport through molecular junctions
is the current for a given bias voltage, given by (in this paper we use atomic units where
$\hbar = e = 1$)
\begin{subequations}
\begin{equation}
	I_L(t) = - \frac{d N_L(t)} {dt} = -\frac{1}{{\rm tr}[\hat{\rho}]} {\rm tr}
        \left\{ \hat{\rho} e^{i\hat{H}t} i[\hat{H}, \hat{N}_{L}] e^{-i\hat{H}t} \right\},
\end{equation}
\begin{equation}
	I_R(t) = \frac{d N_R(t)} {dt} = \frac{1}{{\rm tr}[\hat{\rho}]} {\rm tr}
        \left\{ \hat{\rho} e^{i\hat{H}t} i[\hat{H}, \hat{N}_{R}] e^{-i\hat{H}t} \right\}.
\end{equation}
\end{subequations}
Here, $N_{L/R}(t)$ denotes the  time-dependent charge  in each lead, defined as 
\begin{equation}
	N_{\zeta}(t) = \frac{1}{{\rm tr}[\hat{\rho}]} {\rm tr}
        [\hat{\rho} e^{i\hat{H}t} \hat{N}_{\zeta} e^{-i\hat{H}t} ], \;\;\; \zeta=L, R.
\end{equation}
In the expression above $\hat{N}_{\zeta} = \sum_{k_\zeta} c_{k_\zeta}^+ c_{k_\zeta}$
is the occupation number operator for the electrons in each lead ($\zeta=L, R$) and $\hat{\rho}$ is the 
initial density matrix representing a grand-canonical ensemble for each lead and a certain
preparation for the bridge state
\begin{subequations}\label{Initden}
\begin{equation}
	\hat{\rho} = \hat{\rho}_d^0 \;{\rm exp} \left[ -\beta (\hat{H}_0 
          - \mu_L \hat{N}_L - \mu_R \hat{N}_R) \right],
\end{equation}
\begin{equation}
	\hat{H}_0 = \sum_{k_l} E_{k_l} c_{k_l}^+ c_{k_l}
	+ \sum_{k_r} E_{k_r} c_{k_r}^+ c_{k_r}  + \hat{H}_{\rm nuc}^0.
\end{equation}
\end{subequations}
Here $\hat{\rho}_d^0$ is the initial reduced density matrix for the bridge state, which is usually chosen as
a pure state representing an occupied or an empty bridge state, and $\hat{H}_{\rm nuc}^0$ defines the initial
bath equilibrium distribution, e.g., $\hat{H}_{\rm nuc}$ given above.  The dependence of the steady-state
current on the initial density matrix is a complex issue and is partly addressed in the results section of
this paper.  Finally,
for Hamiltonian (\ref{Htot}) the explicit expression for the current operator is given as
\begin{equation}
	\hat{I}_\zeta \equiv i[\hat{H}, \hat{N}_{\zeta}] = i \sum_{k_\zeta}
	V_{dk_\zeta} ( d^+ c_{k_\zeta} - c_{k_\zeta}^+ d ), \;\;\; 
	\zeta=L, R. 
\end{equation}

The transient behavior of the thus defined currents $I_R(t)$ and $I_L(t)$ is usually different.  
However, the long-time limits of $I_R(t)$ and $I_L(t)$, which define the stationary current, are the same.
It is found that the average current 
\begin{equation}
	I(t) = \frac{1}{2} [ I_R(t) + I_L(t) ],
\end{equation} 
provides better numerical convergence properties by minimizing the transient characteristic, and 
thus will be used in this paper.

Within the model, the left and right leads represent the electronic continuum, each containing 
an infinite number of states.  As mentioned above, in our simulations this
continuous distribution 
is represented by 
a finite number of electronic states.  The number of states required to properly describe the 
continuum limit depends on the time $t$. The situation is thus similar to that of a quantum reactive 
scattering calculation in the presence of a scattering continuum, where, with a finite number of basis 
functions, an appropriate absorbing boundary condition is added to mimic the correct outgoing 
Green's function.\cite{Goldberg1978,Kosloff1986,Neuhauser1989,Seideman1991} 
Employing the same strategy for the present problem, the 
regularized electric current is given by
\begin{equation}
	I^{\rm reg}  = \lim_{\eta \to 0^+} \int_0^{\infty} dt \, \frac{dI(t)}{dt} \, e^{-\eta t}.
\end{equation}

The regularization parameter $\eta$ is similar (though not identical) to the formal convergence parameter
in the definition of  the Green's function in terms of the time evolution operator
\begin{equation}
	G(E^+) = \lim_{\eta \to 0^+} (-i) \int_0^{\infty} dt\, e^{i(E+i\eta-H)t}.
\end{equation}
In numerical calculations, $\eta$ is chosen in a similar way as the absorbing potential used in quantum
scattering calculations.\cite{Goldberg1978,Kosloff1986,Neuhauser1989,Seideman1991} In particular, 
the parameter $\eta$ has to be large enough to accelerate the convergence but still sufficiently small
in order not to affect the correct result.  While in the reactive scattering calculation
$\eta$ is often chosen to be coordinate dependent, in our simulation $\eta$ is chosen
to be time dependent
\begin{equation}\label{damping}
\eta(t) = \left\{
              \begin{array}{ll}
                   0 & \quad (t<\tau)\\
                   \eta_0\cdot (t-\tau)/t & \quad (t>\tau) .
              \end{array}
       \right.
\end{equation}
Here $\eta_0$ is a damping constant, $\tau$ is a cutoff time beyond which a steady state charge
flow is approximately reached.  As the number of electronic states increases, one may choose a 
weaker damping strength $\eta_0$ and/or longer cutoff time $\tau$.  The former approaches zero and the
latter approaches infinity for an infinite number of states.  In practice,
for the systems considered in this work, convergence can be reached with a reasonable number 
of electronic states in the range of 64-128, with a typical $\tau =$ 40-60 fs (a smaller
$\tau$ for less number of states) and $1/\eta_0 =$ 2-5 fs.

To analyze the transport mechanisms, it is also expedient to consider the
population of the electronic state localized on the the molecular bridge,
which is given by
\begin{equation}\label{population}
P_d(t) = \frac{1}{{\rm tr}[\hat{\rho}]} {\rm tr}
        \left\{ \hat{\rho} e^{i\hat{H}t} d^+ d  e^{-i\hat{H}t}
        \right\}.
\end{equation}

\section{The Multilayer Multiconfiguration Time-Dependent Hartree Theory
in Second Quantization Representation}\label{mlsqr}

The accurate treatment of the time-dependent transport problem outlined above requires a 
method that is able to describe the quantum dynamics of a system with many electronic
and nuclear degrees of freedom. To this end, we employ the  recently proposed
Multilayer Multiconfiguration Time-Dependent Hartree Theory
in Second Quantization Representation (ML-MCTDH-SQR).\cite{wan10:78}
This method extends the
ML-MCTDH approach to the treatment of indistinguishable particles.

\subsection{General formulation of the ML-MCTDH theory}

The ML-MCTDH theory\cite{wan03:1289} is a rigorous variational method to propagate wave packets 
in complex systems with many degrees of freedom.  In this approach the wave function is represented by a recursive, layered 
expansion, 
\begin{subequations}\label{psiml}
\begin{equation}\label{L1}
        |\Psi (t) \rangle = \sum_{j_1} \sum_{j_2} ... \sum_{j_p}
        A_{j_1j_2...j_p}(t) \prod_{\kappa=1}^{p}  |\varphi_{j_\kappa}^{(\kappa)} (t) \rangle,
\end{equation}
\begin{equation}\label{L2}
        |\varphi_{j_\kappa}^{(\kappa)}(t)\rangle =  \sum_{i_1} \sum_{i_2} ... \sum_{i_{Q(\kappa)}}
        B_{i_1i_2...i_{Q(\kappa)}}^{\kappa,j_\kappa}(t) \prod_{q=1}^{Q(\kappa)}  
	|v_{i_q}^{(\kappa,q)}(t) \rangle,
\end{equation}
\begin{equation}\label{L3}
        |v_{i_q}^{(\kappa,q)}(t)\rangle  = \sum_{\alpha_1} \sum_{\alpha_2} ... 
	\sum_{\alpha_{M(\kappa,q)}}
        C_{\alpha_1\alpha_2...\alpha_{M(\kappa,q)}}^{\kappa,q,i_q}(t) 
	\prod_{\gamma=1}^{M(\kappa,q)}  
	|\xi_{\alpha_\gamma}^{\kappa,q,\gamma}(t) \rangle,
\end{equation}
\begin{equation}
	... \nonumber
\end{equation}
\end{subequations}
where $A_{j_1j_2...j_p}(t)$, $B_{i_1i_2...i_{Q(\kappa)}}^{\kappa,j_\kappa}(t)$,
$C_{\alpha_1\alpha_2...\alpha_{M(\kappa,q)}}^{\kappa,q,i_q}(t)$ and so on are the
expansion coefficients for the first, second, third, ..., layers, respectively;
$|\varphi_{j_\kappa}^{(\kappa)} (t) \rangle$, $|v_{i_q}^{(\kappa,q)}(t) \rangle$,
$|\xi_{\alpha_\gamma}^{\kappa,q,\gamma}(t) \rangle$, ..., are the ``single particle'' 
functions (SPFs) for the first, second, third, ..., layers.  
In Eq.~(\ref{L1}), $p$ denotes the number of single
particle (SP) groups/subspaces for the first layer.  Similarly, $Q(\kappa)$ in Eq.~(\ref{L2})
is the number of SP groups for the second layer that belongs to the $\kappa$th SP
group in the first layer,  i.e., there are a total of $\sum_{\kappa=1}^{p} Q(\kappa)$
second layer SP groups.  Continuing along the multilayer hierarchy, 
$M(\kappa,q)$ in Eq.~(\ref{L3}) is the number of SP groups for the third layer that belongs 
to the $q$th SP group of the second layer and the $\kappa$th SP group of the first layer,  
resulting in a total of $\sum_{\kappa=1}^{p} \sum_{q=1}^{Q(\kappa)} M(\kappa,q)$ third 
layer SP groups.  Naturally, the size of the  system that the ML-MCTDH theory can treat 
increases with the number of layers in the expansion.  In principle, such a recursive 
expansion can be carried out to an arbitrary number of layers.  The multilayer hierarchy 
is terminated at a particular level by expanding the SPFs in the deepest layer in terms of 
time-independent configurations, each of which may contain several Cartesian degrees of 
freedom.

The variational parameters within the ML-MCTDH theoretical framework are dynamically 
optimized through the use of Dirac-Frenkel variational principle\cite{Frenkel34}
\begin{equation}
        \langle \delta\Psi(t) | i \frac{\partial} {\partial t} - \hat{H} |
        \Psi(t) \rangle = 0,
\end{equation}
which results in a set of coupled, nonlinear differential equations
\begin{subequations}\label{mleqs}
\begin{equation}\label{Aeq2}
        i | \dot{\Psi}(t) \rangle_{\rm L1\; coefficients}
        =  \hat{H}(t) |  \Psi(t)\rangle,
\end{equation}
\begin{equation}\label{Beq2}
        i |\underline{\dot{\varphi}}^k(t) \rangle_{\rm L2\; coefficients} 
	=  [1-\hat{P}^{(\kappa)}(t) ]
        [\hat{\rho}^{(\kappa)}(t) ]^{-1} \langle \hat{H} \rangle^{(\kappa)} (t)
        |\underline{\varphi}^{(\kappa)}(t)\rangle,
\end{equation}
\begin{equation}\label{Ceq2}
        i |\underline{\dot{v}}^{(\kappa,q)}(t) \rangle_{\rm L3\; coefficients}  
	=  [1-\hat{P}^{(\kappa,q)}_{\rm L2}(t) ]
        [\hat{\varrho}^{(\kappa,q)}_{\rm L2} (t)]^{-1} \langle \hat{\mathcal{H}} 
	\rangle^{(\kappa,q)}_{\rm L2} (t) |\underline{v}^{(\kappa,q)}(t) \rangle,
\end{equation}
\begin{equation}\label{Deq2}
        i |\underline{\dot{\xi}}^{(\kappa,q,\gamma)}(t) \rangle_{\rm L4\; coefficients}  
	=  [1-\hat{P}^{(\kappa,q,\gamma)}_{\rm L3}(t) ]
        [\hat{\varrho}^{(\kappa,q,\gamma)}_{\rm L3}(t)]^{-1} \langle \hat{\mathcal{H}} 
	\rangle^{(\kappa,q,\gamma)}_{\rm L3} (t) |\underline{\xi}^{(\kappa,q,\gamma)}(t) \rangle,
\end{equation}
\begin{equation}
	... \nonumber
\end{equation}
\end{subequations}
For referencing purpose we label the SP subspaces from top to bottom layers as level one 
(L1), level 2 (L2), and so on.  In the notation used in Eq.\ (\ref{mleqs}) the time 
derivatives on the left hand side (represented by a dot) are only performed with respect 
to the expansion coefficients of a particular layer (denoted by the respective 
subscript).  For example, the time derivative in Eq.~(\ref{Aeq2}) acts only on the L1 
expansion coefficient $A_{j_1j_2...j_p}(t)$; the time derivative in Eq.~(\ref{Beq2}) is 
on the L2 expansion coefficient $B_{i_1i_2...i_{Q(\kappa)}}^{\kappa,j_\kappa}(t)$; etc.  
In our convention, for a $N$-layer version of the ML-MCTDH theory there are $N+1$ levels 
of expansion coefficients.  In this sense the conventional wave packet propagation method 
is a ``zero-layer'' MCTDH approach.

In practical implementations various intermediate quantities are defined within the subspaces
of each layer of the wave function.\cite{wan03:1289}  For example, the top-layer Hamiltonian
matrix $[\hat{H}(t)]_{JL} \equiv \langle \Phi_J(t) | \hat{H} | \Phi_L(t) \rangle$,
where $|\Phi_J (t) \rangle \equiv  \prod_{\kappa=1}^{p}  |\varphi_{j_\kappa}^{(\kappa)} (t) \rangle$
is a configuration in the L1 subspace; the reduced density matrices
$\hat{\rho}^{(\kappa)}(t)$, $\hat{\varrho}^{(\kappa,q)}_{\rm L2} (t)$, 
$\hat{\varrho}^{(\kappa,q,\gamma)}_{\rm L3}(t)$, ..., 
$\hat{\varrho}^{(\kappa,q,\gamma,...)}_{\rm LN}(t)$ for the
first, second, third, and Nth layers, respectively; and the 
corresponding mean-field operators $\langle \hat{H} \rangle^{(\kappa)} (t)$,
$\langle \hat{\mathcal{H}} \rangle^{(\kappa,q)}_{\rm L2} (t)$, 
$\langle \hat{\mathcal{H}} \rangle^{(\kappa,q,\gamma)}_{\rm L3} (t)$, ...,
$\langle \hat{\mathcal{H}} \rangle^{(\kappa,q,\gamma,...)}_{\rm LN} (t)$.
These operators can be recursively evaluated by means of the
single hole functions $|\Psi_m^{(\kappa)}(t) \rangle$, 
$|g_{{\rm L2};\;m,s}^{(\kappa,q)}(t) \rangle$, 
$|g_{{\rm L3};\;m,s}^{(\kappa,q,\gamma)}(t) \rangle$, ..., for the first, second, third, and
further layers.\cite{wan03:1289,wan10:78,wan09:024114}

The introduction of this recursive, dynamically optimized layering scheme in the ML-MCTDH 
wavefunction provides more flexibility in the variational functional, which results in 
tremendous gain in our ability to study large quantum many-body systems.  During the past 
few years, significant progress has been made in further development of the theory to simulate 
quantum dynamics and nonlinear spectroscopy of ultrafast  electron transfer reactions in 
condensed phases.\cite{tho06:210,wan06:034114,kon06:1364,wan07:10369,kon07:11970,tho07:153313,wan08:139,wan08:115005,ego08:214303,vel09:094109,wan10:78,Vendrell11} 
The theory has also been generalized to study heat transfer through molecular 
junctions\cite{vel08:325} and to calculate rate constants for model proton transfer reactions 
in molecules in solution.\cite{wan06:174502,cra07:144503}
Recent work of Manthe has introduced an even more adaptive formulation based on
a layered correlation discrete variable representation (CDVR).\cite{man08:164116,man09:054109}

\subsection{Treating identical particles using the second quantization representation of Fock space}

Despite its previous success, the original ML-MCTDH theory was not directly applicable to
studying systems of identical quantum particles.  This is because an ordinary Hartree product in 
the first quantized picture is only suitable to describe a configuration for a system of 
distinguishable particles.  To handle systems of identical particles, one strategy is to 
employ a properly symmetrized wave function, i.e., permanents in a bosonic case or Slater 
determinants in a fermionic case.  This led to the MCTDHF 
approach\cite{kat04:533,cai05:012712,nes05:124102} for treating identical fermions and the MCTDHB 
approach\cite{alo08:033613} for treating identical bosons as well as combinations 
thereof.\cite{alo07:154103}  However, this wave function-based symmetrization is only applicable 
to the single layer MCTDH theory but is incompatible with the ML-MCTDH theory with more layers 
--- there is no obvious analog of a multilayer Hartree configuration if permanents/determinants 
are used to represent the wave function.  As a result, the ability to treat much larger quantum 
systems numerically exactly was severely limited.

To overcome this limitation we proposed a novel approach\cite{wan09:024114} that follows a 
fundamentally different route to tackle quantum dynamics of indistinguishable particles --- 
an operator-based method that employs the second quantization formalism of many-particle quantum 
theory.  Thereby the variation is carried out entirely in the abstract Fock space using the 
occupation number representation.  This differs from many previous methods where the second 
quantization formalism is only used as a convenient tool to derive intermediate expressions for 
the first quantized form.  In the new approach the burden of handling symmetries of identical 
particles in a numerical variational calculation is shifted completely from wave functions 
to the algebraic properties of operators.

The procedure can be illustrated by considering a system of identical fermions.
In the first quantized representation a Slater determinant 
$|\chi_{P_1}\chi_{P_2}...\chi_{P_N}\rangle$ 
describes an anti-symmetric $N$-particle state by choosing $N$ spin orbitals 
out of the $M$ orthonormal spin-adapted basis functions,  
$\{ \chi_1({\bf x}), \chi_2({\bf x}), ..., \chi_M({\bf x}) \}$. 
All possible Slater determinants in this form constitute the fermionic subspace of 
the $N$-particle Hilbert space, denoted by ${\cal H}(M,N)$.
The Fock space ${\cal F}$(M) is formed by considering an arbitrary number of particles
\begin{equation}\label{decomp}
	{\cal F}(M) = {\cal H}(M,0) \oplus {\cal H}(M,1)  \oplus {\cal H}(M,2) 
	\oplus ... \oplus {\cal H}(M,M).
\end{equation}
In second quantization a convenient basis to represent the Fock space is the occupation 
number basis
\begin{equation}\label{ONvec}
	|{\bf n}\rangle \equiv |n_1, n_2, ..., n_M\rangle,
\end{equation}
where  $n_P$ can be either 1 if the one-particle state $\chi_P$ is occupied
[i.e., present in the original Slater determinant]  or 0 if it is unoccupied.
The number of particles in state $|{\bf n}\rangle$ is given by
$N = \sum_{P=1}^M n_P$,
and thus ${\cal H}(M,N)$ contains all occupation-number vectors with $N$ particles.
Each occupation-number state is defined by acting
a series of creation operators $a_P^+$ on the vacuum state,
$|{\bf n}\rangle = \prod_{P=1}^M (a_P^+)^{n_P} |{\rm vac}\rangle.$

In contrast to a Slater determinant, the occupation-number state
$|{\bf n}\rangle$ can be formally written in the form of a Hartree product,\cite{Greiner98}
\begin{equation}
	|{\bf n}\rangle = \prod_{P=1}^M |n_P\rangle
 	\equiv |n_1\rangle |n_2\rangle ... |n_M\rangle.
\end{equation}
This formal factorization suggests a different decomposition of the Fock space 
\begin{subequations}
\begin{equation}\label{decomp2}
	{\cal F}(M) = f_1(1) \otimes f_2(1) \otimes ... \otimes f_\kappa(1) \otimes ... \otimes
		f_M(1),
\end{equation}
where $f_\kappa(1)$ represents a single spin-orbital subspace with two possibilities/states: 
occupied or unoccupied.  Bigger subspaces can be formed by grouping a few spin orbitals 
together
\begin{equation}\label{decomp2b}
	{\cal F}(M) = {f}_1(m_1) \otimes {f}_2(m_2) \otimes ... 
\otimes {f}_\kappa(m_\kappa) \otimes ... \otimes
		{f}_L(m_L),  \hspace{1cm} M = \sum_{\kappa=1}^{L} m_\kappa,
\end{equation}
\end{subequations}
where $f_\kappa(m_\kappa)$ represents the subspace with $m_\kappa$ spin orbitals and thus 
$2^{m_\kappa}$ states.

The decomposition schemes in Eqs.~(\ref{decomp2}) and (\ref{decomp2b}) are conceptually 
different from that in (\ref{decomp}).  
They no longer require dealing with a full-length occupation-number vector in one step
and treating it as a whole, unbreakable object like the original Slater determinant, 
but rather focus on each subspace 
$f_\kappa(m_\kappa)$ containing $2^{m_\kappa}$ linearly independent sub-vectors 
$\{\phi^{(\kappa)}_{I_\kappa}\}$, $I_\kappa=1,...,2^{m_\kappa}$.  The crucial point is 
that in a variational calculation each sub-vector in the $\kappa$th subspace is not restricted 
to a particular fixed basis vector $\{\phi^{(\kappa)}_{I_\kappa}\}$, but may also be an 
appropriate superposition of all the sub-vectors in this subspace.  One may then express the
overall wave function in the same multilayer form as in Eq.~(\ref{psiml}).  Take, for example,
three explicit layers in (\ref{psiml}), the deepest layer (3rd layer here) is expanded in the 
full basis sub-vectors in the Fock subspace as,
\begin{equation}
	|\xi_{\alpha_\gamma}^{\kappa,q,\gamma}(t) \rangle =
	\sum_{n_1=0}^1 \sum_{n_2=0}^1 ... \sum_{n_{m(\kappa,q,\gamma)}=0}^1
        D_{n_1n_2...n_{m(\kappa,q,\gamma)}}^{\kappa,q,\gamma,\alpha_\gamma}(t)\; |n_1\rangle |n_2\rangle ...  
	|n_{m(\kappa,q,\gamma)} \rangle.
\end{equation}
$|\Psi (t) \rangle$ in Eq.~(\ref{psiml}) is then built ``bottom-up'' from the optimal, 
time-dependent SPFs for the subspaces.

After introducing the occupation-number representation of the Fock space, 
the Hamiltonian can be expressed in terms of fermionic creation/annihilation operators via 
the standard procedure in the second quantization
formalism.\cite{Fetter75,Mahan81}  The overall method is 
thus a ML-MCTDH theory in second quantization representation (SQR).
The major difference between the ML-MCTDH-SQR theory for identical fermions and the previous 
ML-MCTDH theory for distinguishable particles is the way how operators act. In the second 
quantized form the fermionic creation/annihilation operators fulfill the 
anti-commutation relations
\begin{equation}\label{anticomm}
	\{ a_P, a_Q^+ \} \equiv a_P a_Q^+ + a_Q^+ a_P = \delta_{PQ},
	\hspace{1cm}
	\{ a_P^+, a_Q^+ \} = \{ a_P, a_Q \} = 0.
\end{equation}
The symmetry of identical particles is thus realized by enforcing such algebraic
properties of the operators.

The practical procedure can be illustrated by considering
a single layer theory in the form of Eq.~(\ref{decomp2b}), where each
SP group $\kappa$ corresponds to a Fock subspace in (\ref{decomp2b})
\begin{subequations}
\begin{equation}
\label{mcsqr}
        |\Psi (t) \rangle 
	= \sum_{j_1} \sum_{j_2} ... \sum_{j_L}
        A_{j_1j_2...j_L}(t) \prod_{\kappa=1}^{L}  |\varphi_{j_\kappa}^{(\kappa)} (t) \rangle,
\end{equation}
\begin{equation}\label{spsqr}
        |\varphi_{j_\kappa}^{(\kappa)}(t)\rangle = \sum_{I_\kappa=1}^{2^{m_\kappa}}
	B_{I_\kappa}^{\kappa,j_\kappa}(t) |\phi^{(\kappa)}_{I_\kappa} \rangle \equiv
	\sum_{n_1=0}^1 \sum_{n_2=0}^1 ... \sum_{n_{m_\kappa}=0}^1
        B_{n_1n_2...n_{m_\kappa}}^{\kappa,j_\kappa}(t)\; |n_1\rangle |n_2\rangle ...  
	|n_{m_\kappa} \rangle.
\end{equation}
\end{subequations}
Without loss of generality let us consider acting a creation
operator $(a_{\nu}^{(\kappa)})^+$ on the SPFs. 
In practical implementation this operation is equivalent to
\begin{equation}\label{fermcreat}
	({a}_{\nu}^{(\kappa)})^+ = \left( \prod_{\mu=1}^{\kappa-1}\; \hat{S}_\mu \right)
	\; ({\tilde{a}}_{\nu}^{(\kappa)})^+,
\end{equation}
where $\hat{S}_\mu$ ($\mu=1,2,...,\kappa-1$) is the permutation sign operator that accounts 
for permuting $({a}_{\nu}^{(\kappa)})^+$ from the first subspace all the way through to 
the $\kappa$th subspace, and
$({\tilde{a}}_{\nu}^{(\kappa)})^+$ is the reduced creation operator that only takes care of 
the fermionic anti-commutation relation in the $\kappa$th subspace.  The operator-based
anti-commutation constraint (\ref{anticomm}) results in the following operations
\begin{subequations}\label{permutesign}
\begin{equation}
	({\tilde{a}}_{\nu}^{(\kappa)})^+  |\varphi_{j_\kappa}^{(\kappa)}(t)\rangle =
	\sum_{n_1=0}^1 \sum_{n_2=0}^1 ... \sum_{n_{m_\kappa}=0}^1 \; \delta_{n_\nu,0}
	\left[\prod_{q=1}^{\nu-1} (-1)^{n_q}\right]\;  B_{n_1n_2...n_{m_\kappa}}^{\kappa,j_\kappa}(t)\;
	|n_1\rangle |n_2\rangle ... |1_\nu\rangle ... |n_{m_\kappa} \rangle,
\end{equation}
\begin{equation}
	\hat{S}_\mu |\varphi_{j_\mu}^{(\mu)}(t)\rangle =
	\sum_{n_1=0}^1 \sum_{n_2=0}^1 ... \sum_{n_{m_\mu}=0}^1 \;
	\left[\prod_{q=1}^{m_\mu} (-1)^{n_q}\right]\;  B_{n_1n_2...n_{m_\mu}}^{\mu,j_\mu}(t)\;
	|n_1\rangle |n_2\rangle ... |n_{m_\mu} \rangle.
\end{equation}
\end{subequations}
I.e.,  $({\tilde{a}}_{\nu}^{(\kappa)})^+$ not only creates a particle in the 
$\nu$th spin orbital if it is vacant, but also affects the sign of each term in 
this SPF according to where 
$\nu$ is located and what the occupations are prior to it.  Furthermore, the permutation sign 
operators  $\hat{S}_\mu$, $\mu=1,2,...,\kappa-1$, incorporate the sign changes of the remaining 
spin orbitals in all the SPFs whose subspaces are prior to that of 
$(\tilde{a}_{\nu}^{(\kappa)})^+$.

The implementation of Eq.~(\ref{fermcreat}) is sophisticated but nevertheless a routine practice
in the MCTDH or ML-MCTDH theory --- a product of operators.  Thereby, the action of each 
Hamiltonian term (product of creation/annihilation operators) can be split into a series of 
operations on individual Fock subspaces.

The generalization from the single layer to the multilayer case is tedious but straightforward. 
The Fock space is decomposed in a recursive, layered fashion --- the spin orbitals here in the 
ML-MCTDH-SQR theory are treated in the same way as the degrees of freedom in the original 
ML-MCTDH theory, except that the orderings of all the SP groups in all layers need to be recorded 
and maintained in later manipulations.  The wave function can then be recursively expanded via 
Eq.~(\ref{psiml}).  More importantly, the equations of motion have the same form as in the original 
ML-MCTDH theory.  The only difference is that each creation/annihilation operator of the 
Hamiltonian is effectively a product of operators: a reduced creation/annihilation operator 
that only acts on the bottom-layer SPFs for the Fock subspace it belongs to, and a series of 
permutation sign operators that accounts for the fermionic anti-commutation relations of all 
the spin orbitals prior to it.

In the second quantized form,  the wave function is represented in the abstract Fock space 
employing the occupation number basis.  As a result, it can be expanded in the same multilayer 
form as that for systems of distinguishable particles.  It is thus possible to extend the 
numerically exact treatment to much larger systems.  The symmetry of the wave function in 
the first quantized form is shifted to the operator algebra in the second quantized form.  
The key point is that, for both phenomenological models and more fundamental theories, 
there are only a limited number of combination of fundamental operators.  
For example, in electronic structure theory only one- and two-electron operators are present.  
This means that one never needs to handle all, redundant possibilities of operator combinations as 
offered by the determinant form in the first quantized framework.  It is exactly this property 
that provides the flexibility of representing the wave functions in multilayer form and treat 
them accurately and efficiently within the ML-MCTDH-SQR theory. It is also noted that 
the ML-MCTDH-SQR approach outlined above for fermions has
also be formulated for bosons or combinations of fermions, bosons and
distinguishable particles.\cite{wan09:024114} Here, we apply it to vibrationally coupled electron transport, which involves a combination of vibrational degrees of freedom and indistinguishable electrons.

\section{Results and Discussion}\label{results}

In this section we present applications of the ML-MCTDH-SQR methodology to
vibrationally coupled electron transport employing the model described in
Sec.\ \ref{modeltight}. We first consider a model with the following set of 
electronic parameters: 
The energy of the discrete state 
$E_d$ is located 0.5 eV above the Fermi energy of the leads ($E_f=0$).  
The tight-binding parameters for the function $\Gamma (E)$
are $\alpha_e = 0.2$ eV, $\beta_e = 1$ eV, corresponding to a moderate
molecule-lead coupling an a bandwidth of 4 eV.

\begin{figure}[!ht]
\begin{flushleft}
(a)
\end{flushleft}
\includegraphics[clip,width=0.45\textwidth]{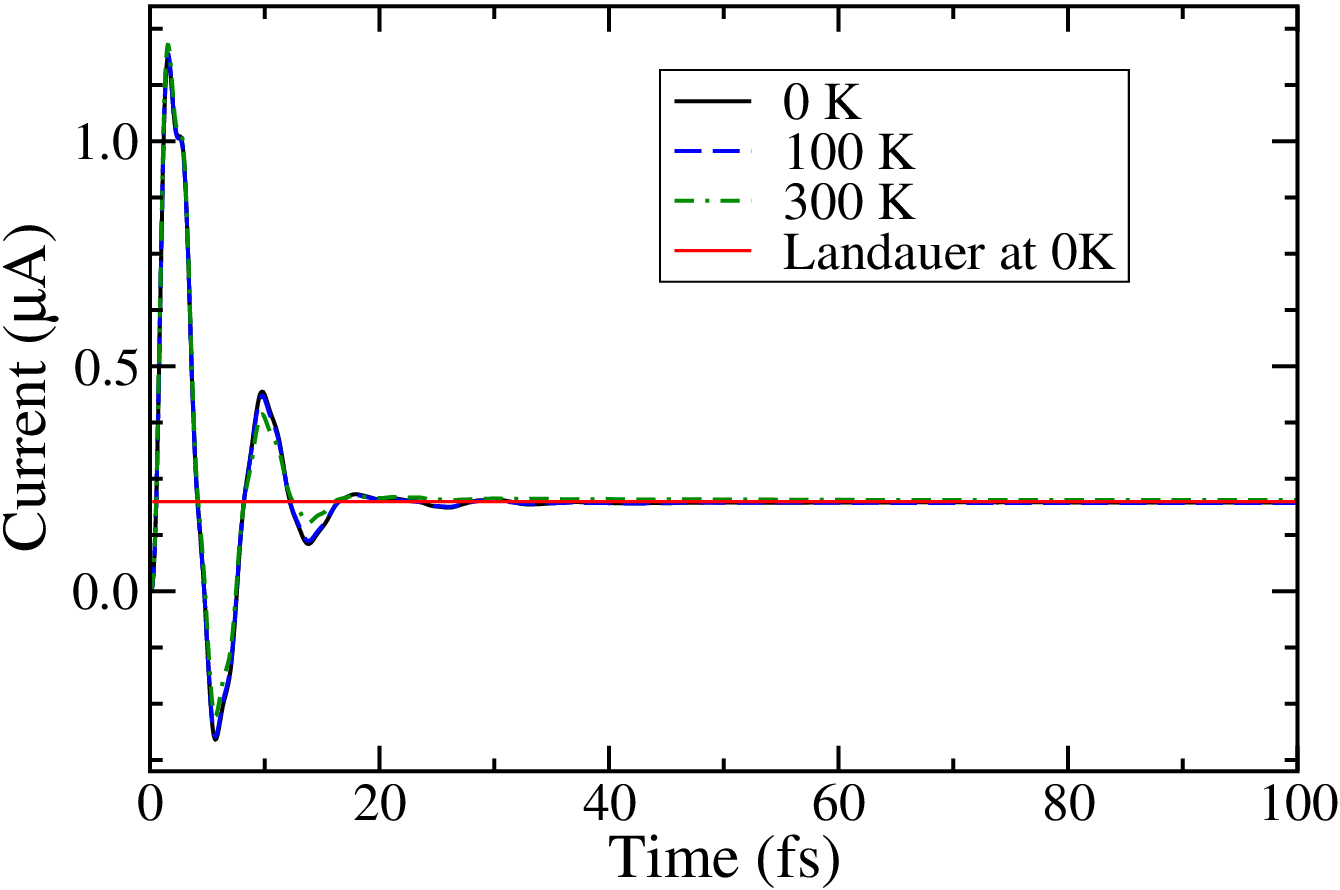}

\begin{flushleft}
(b)
\end{flushleft}
\includegraphics[clip,width=0.45\textwidth]{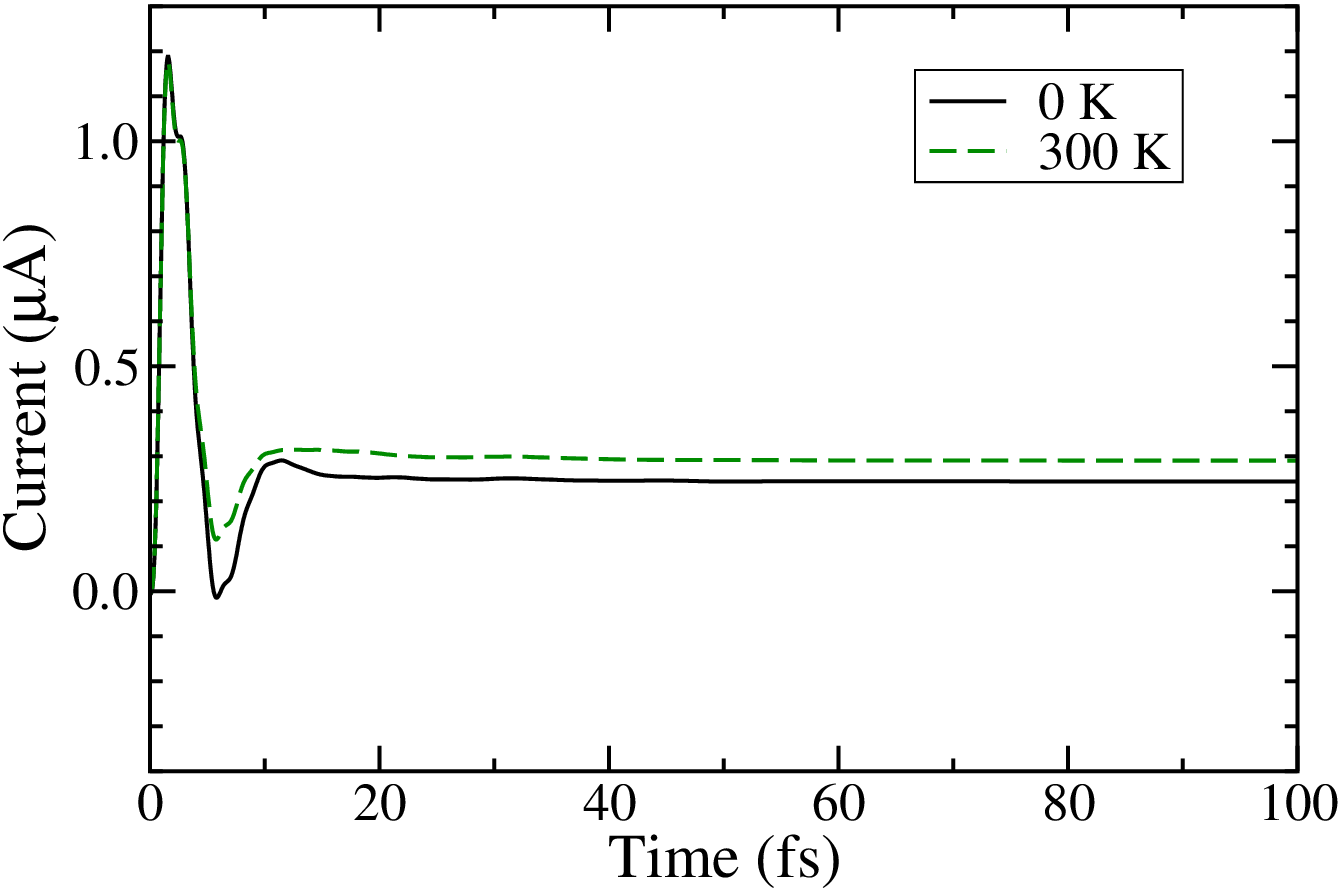}

\caption{Time-dependent current $I(t)$ at different temperatures.
The parameters are: $\alpha_e = 0.2$eV, $\beta_e = 1$eV, $E_d - E_f = 0.5$eV,
and $V = 0.2$V. Results in panel (a) have been obtained without coupling to
the vibrations ($\lambda =0$).  In panel  
(b) couplings to an Ohmic bath of vibrational modes with
parameters are $\lambda = 2000 {\rm cm}^{-1}$ and $\omega_c = 500{\rm
  cm}^{-1}$ is included. The red line in panel (a) shows the
current for a
purely electronic model ($\lambda = 0$) as obtained from Landauer theory}
\label{fig1}
\end{figure}

Figure~\ref{fig1} shows the time-dependent current for low bias voltage, $V=0.2$ V,
and  a range of different temperatures, 0 - 300 K.
Panel (a) depicts the purely electronic current obtained without coupling to
the vibrational degrees of freedom ($\alpha = 0$).
In this case significant electronic coherence is 
observed for the current $I(t)$ at short time, which decreases for longer time.  
A plateau of $I(t)$ is reached in relative short time ($\sim 30$ fs), 
which demonstrates the feasibility of using a time-dependent approach to obtain the 
stationary current. In contrast to the transient characteristics, 
for the purely electronic problem considered in Fig.~\ref{fig1}(a) the stationary  
current can also be obtained exactly from the Green's function or the scattering
theory approach, employing e.g. Landauer
theory.\cite{lan57:223,Datta95,Cuevas10} 
The thus obtained stationary value of the current agrees with the simulation result.
The results in Fig.~\ref{fig1}(a) also illustrate the fairly weak temperature dependence of 
$I(t)$ for this set of electronic parameters.

Figure~\ref{fig1}(b) shows results for the same electronic parameters as in Fig.~\ref{fig1}(a), 
however including the coupling to the vibrational bath.   The characteristic
frequency of the bath has been chosen as
$\omega_c = 500$~cm$^{-1}$ and the overall electronic-nuclear
coupling strength is determined by the reorganization energy, 
$\lambda = 2\alpha\omega_c = 2000$~cm$^{-1}$. These parameters represent
typical values for polyatomic molecules.\cite{Benesch08} It is noted that in contrast to 
many previous treatments of vibrationally coupled electron transport in molecular junctions, 
the present method allows a nonperturbative, in principle numerically exact treatment of 
this nonequilibrium problem.   The results show that
the inclusion of electronic-vibrational coupling has a significant effect on
the transport characteristics.
In particular, it causes a quenching of the electronic coherence.  As a result, 
the time scale on which the current $I(t)$ reaches its stationary value is
shorter.  
Furthermore, the temperature dependence of the current is more pronounced than in the 
corresponding purely electronic case.  It is also noted that in this particular physical regime the 
value of the current is larger than for purely electronic transport ({\em vide infra}).

\begin{figure}[!h]

\includegraphics[clip,width=0.45\textwidth]{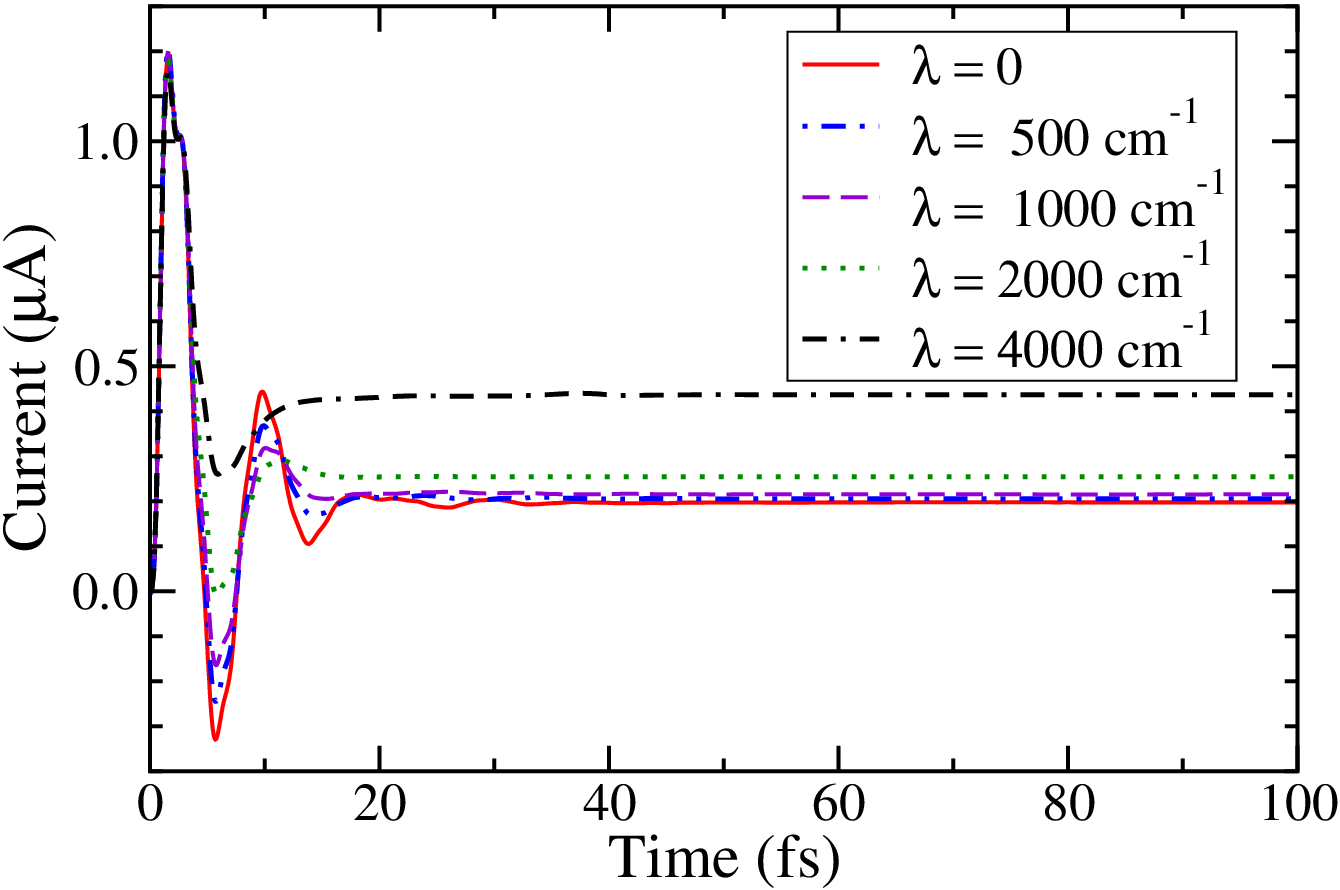}

\caption{Time-dependent current $I(t)$ for different coupling strengths to the vibrational
  bath as specified by the reorganization energy $\lambda$
at temperature $T = 0$ and bias voltage $V=0.2$V.  All other parameters are
the same as in Fig.~\ref{fig1}(b). }
\label{fig2}
\end{figure}

Figure~\ref{fig2} shows the time-dependent current for different
electronic-vibrational coupling strengths.  
While smaller electronic-vibrational coupling 
($\lambda \leq 1000$ ${\rm cm}^{-1}$) causes mostly decoherence in the transient regime, 
larger coupling is seen to influence also the stationary value of the current 
significantly.  This can be understood qualitatively from the fact that the
coupling to the vibrations effectively lowers the energy level of the bridge
(polaron shift). For the given 
voltage, the bare energy of the electronic bridge state is still outside the
conductance window, which is defined by the chemical potentials of the two electrodes. The
coupling to the vibrations brings the level closer to the chemical potential of the electrodes. 
As a result, the current is enhanced. 
The value of this polaron shift of the energy is given  by the
reorganization energy $\lambda$. For example, while  for a value of  $\lambda =  1000$ ${\rm
  cm}^{-1}$ the predominant transport mechanism is nonresonant tunneling, 
for a value of of  $\lambda =  4000$ ${\rm
  cm}^{-1}$ the polaron-shifted bridge state is already inside the conductance
window between the chemical potential
of the electrodes and thus the transport mechanism is resonant tunneling. 

\begin{figure}[!h]

\includegraphics[clip,width=0.45\textwidth]{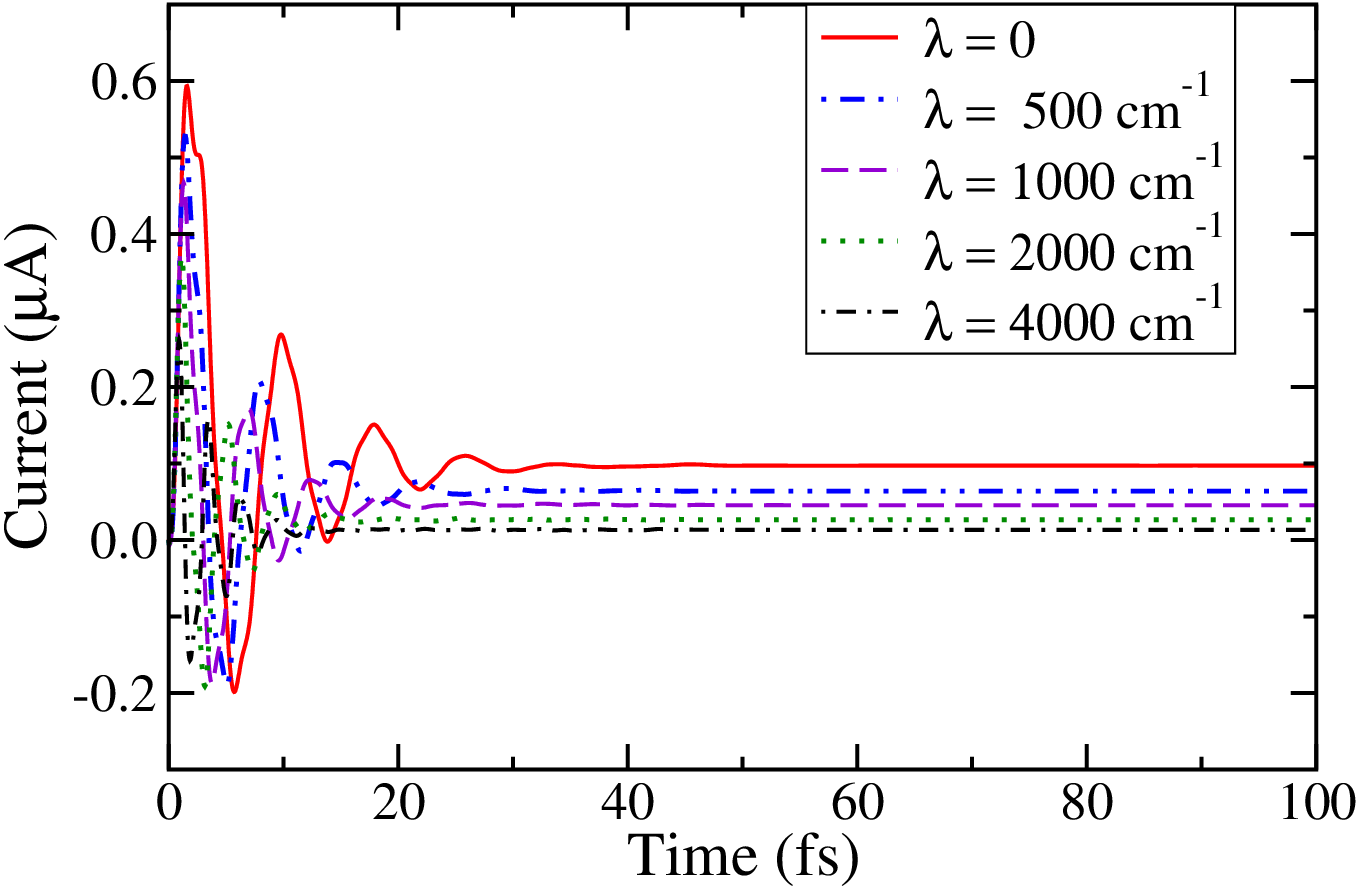}

\caption{Time-dependent current $I(t)$ for different coupling strengths to the vibrational
  bath as specified by the reorganization energy $\lambda$ 
at temperature $T = 0$ and bias voltage $V=0.1$V.  
Except for $E_d - E_f = -0.5$eV, all other parameters are the 
same as in Fig.~\ref{fig1}(b). }
\label{fig3}
\end{figure}

\begin{figure}[!ht]
\begin{flushleft}
(a)
\end{flushleft}
\includegraphics[clip,width=0.45\textwidth]{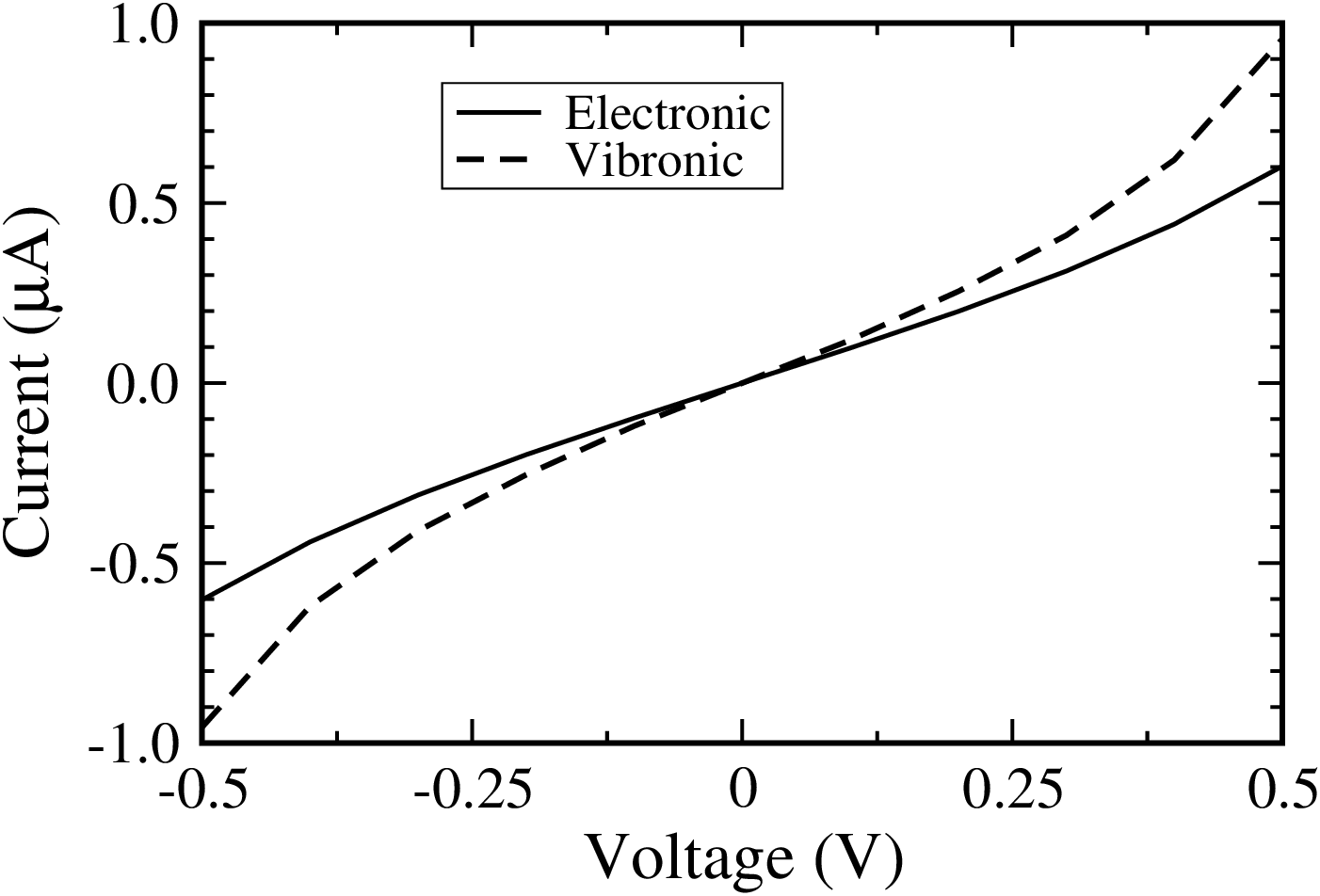}

\begin{flushleft}
(b)
\end{flushleft}
\includegraphics[clip,width=0.45\textwidth]{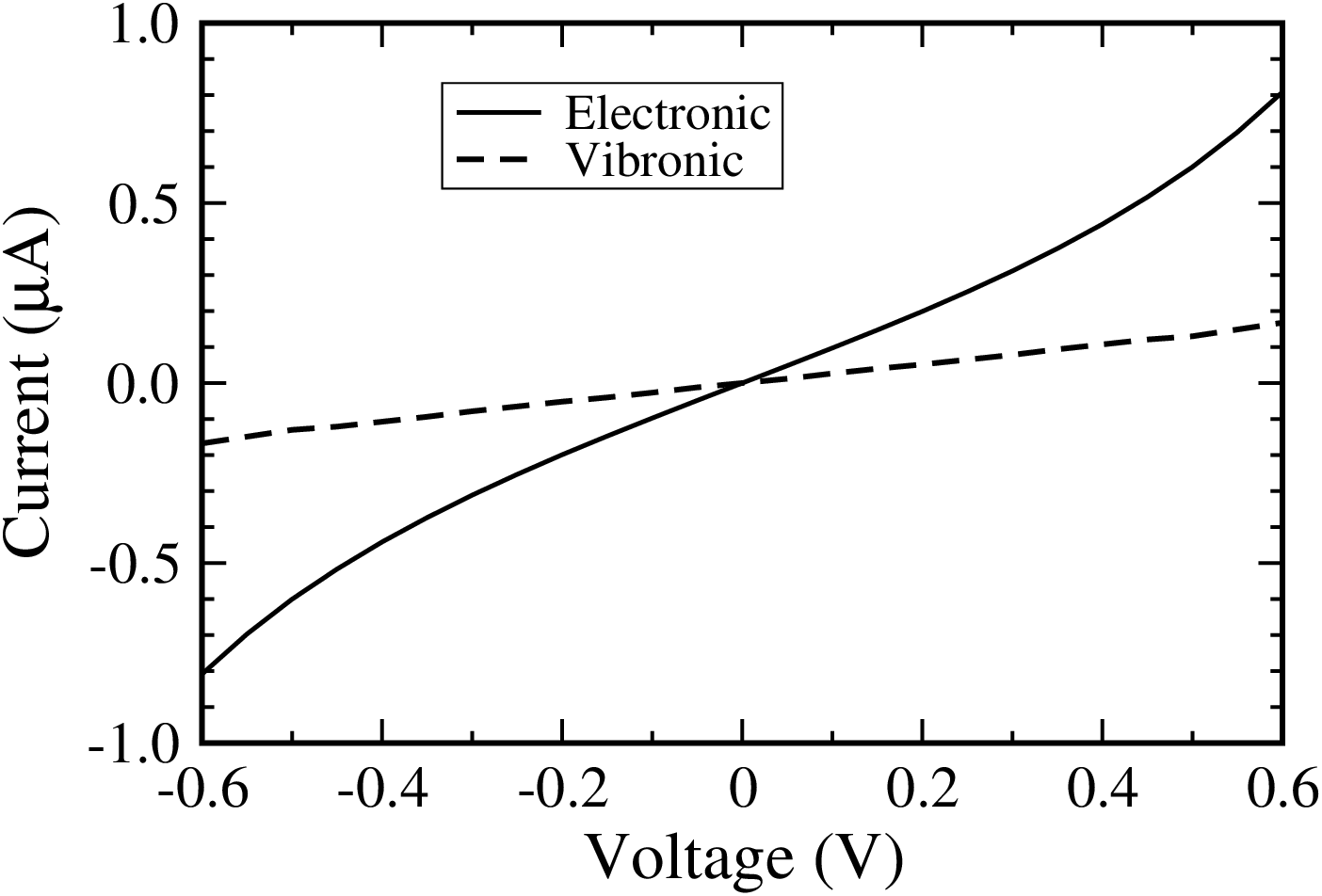}

\caption{Current-voltage characteristics for vibrational coupled electron transport.
The parameters for the electronic lead states are $\alpha_e = 0.2$eV
and $\beta_e = 1$eV.
The vibrational parameters are $\lambda = 2000 {\rm cm}^{-1}$ and $\omega_c = 500{\rm cm}^{-1}$.
The energy of the bridge state is located at:
(a) $E_d - E_f = 0.5$eV, (b) $E_d - E_f = -0.5$eV. The dashed lines depict the
current for a
purely electronic model ($\lambda = 0$)}
\label{fig4}
\end{figure}

We next consider a model with the same parameters except that the energy of
the bridge state is below the Fermi energy, $E_d - E_f= -0.5$ eV.  
As shown in Figure~\ref{fig3}, in this case an increase in the electronic-vibrational
coupling strength not only quenches the electronic coherence but also reduces the
current monotonically. This is due to the fact that if the  bridge state
is located  below the Fermi levels of the leads (Figure~\ref{fig3}) 
the coupling to the vibrational bath will shift its effective energy further
away from  the resonant transport
regime. Note that, as implied by the model Hamiltonian, Eq.\ (\ref{Htot}), the
calculations  depicted in Figure~\ref{fig3}
use the same electronic reference state as for Figure~\ref{fig2}, i.e. the
polaron shift is to lower energies for both calculations.
Figure~\ref{fig4}
displays the current-voltage characteristics for the two models with the same electronic
parameters as in Figs.\ \ref{fig3} and \ref{fig4}, respectively, coupled to a
bath with a characteristic frequency of $\omega_c = 500$~cm$^{-1}$ and a reorganization 
energy of $\lambda = 2000$~cm$^{-1}$. 
The simulation results are compared
with the results for a purely electronic system obtained using the Landauer
formula. The results show a pronounced influence of the vibrational coupling.  In this
particular parameter regime, for $E_d - E_f= 0.5$ eV the vibrationally coupled transport 
current is higher than that for the purely electronic model, whereas for $E_d - E_f= -0.5$ eV
the situation is opposite.  Although this can be qualitatively explained by the polaron 
shift of the energy of the bridge state as discussed above, the actual
quantitative prediction  is more complex and
requires a non-perturbative, accurate approach ({\em vide infra}).

It is worthwhile to point out that the initial condition used in the
expression for the current, Eq.~(\ref{Initden}), is not unique.
For example, one may choose an initially unoccupied bridge state and an unshifted bath of
oscillators, i.e.\ $H_{\rm nuc}$ as given in Eq.~(\ref{Htot}). On the other hand, one may also start with
a fully occupied bridge state and a bath of oscillators in equilibrium with the occupied 
bridge state
\begin{equation}
        H_{\rm nuc}' = \frac{1}{2} \sum_j \left[ P_j^2 + \omega_j^2 \left(Q_j + 
	\frac{c_j}{\omega_j^2}\right)^2 \right].
\end{equation}
Other initial states may also be prepared. 
Thus the question arises, whether
the stationary current depends on the initial state that is used in the
time-dependent simulation.

\begin{figure}[!ht]
\begin{flushleft}
(a)
\end{flushleft}
\includegraphics[clip,width=0.45\textwidth]{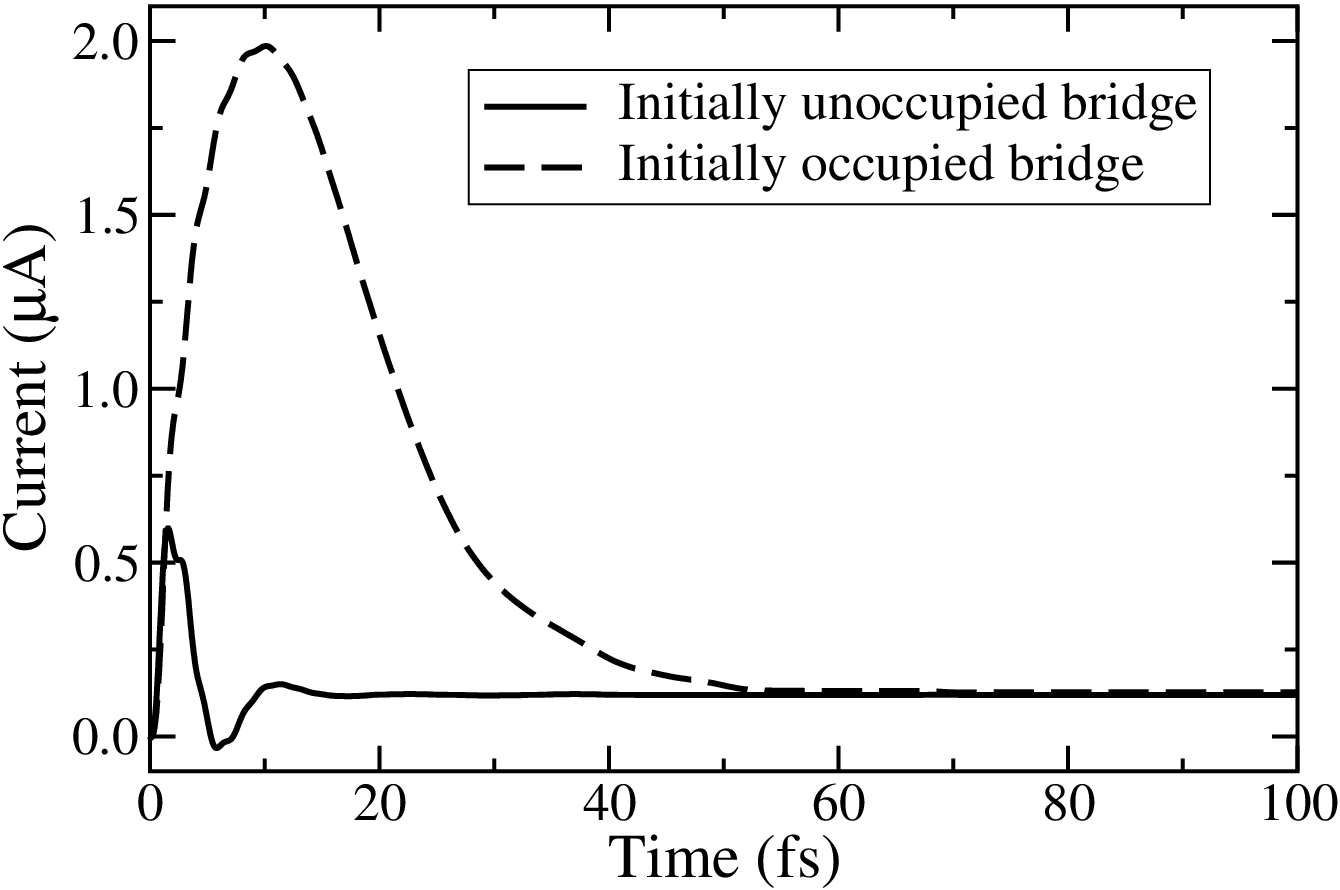}

\begin{flushleft}
(b)
\end{flushleft}
\includegraphics[clip,width=0.45\textwidth]{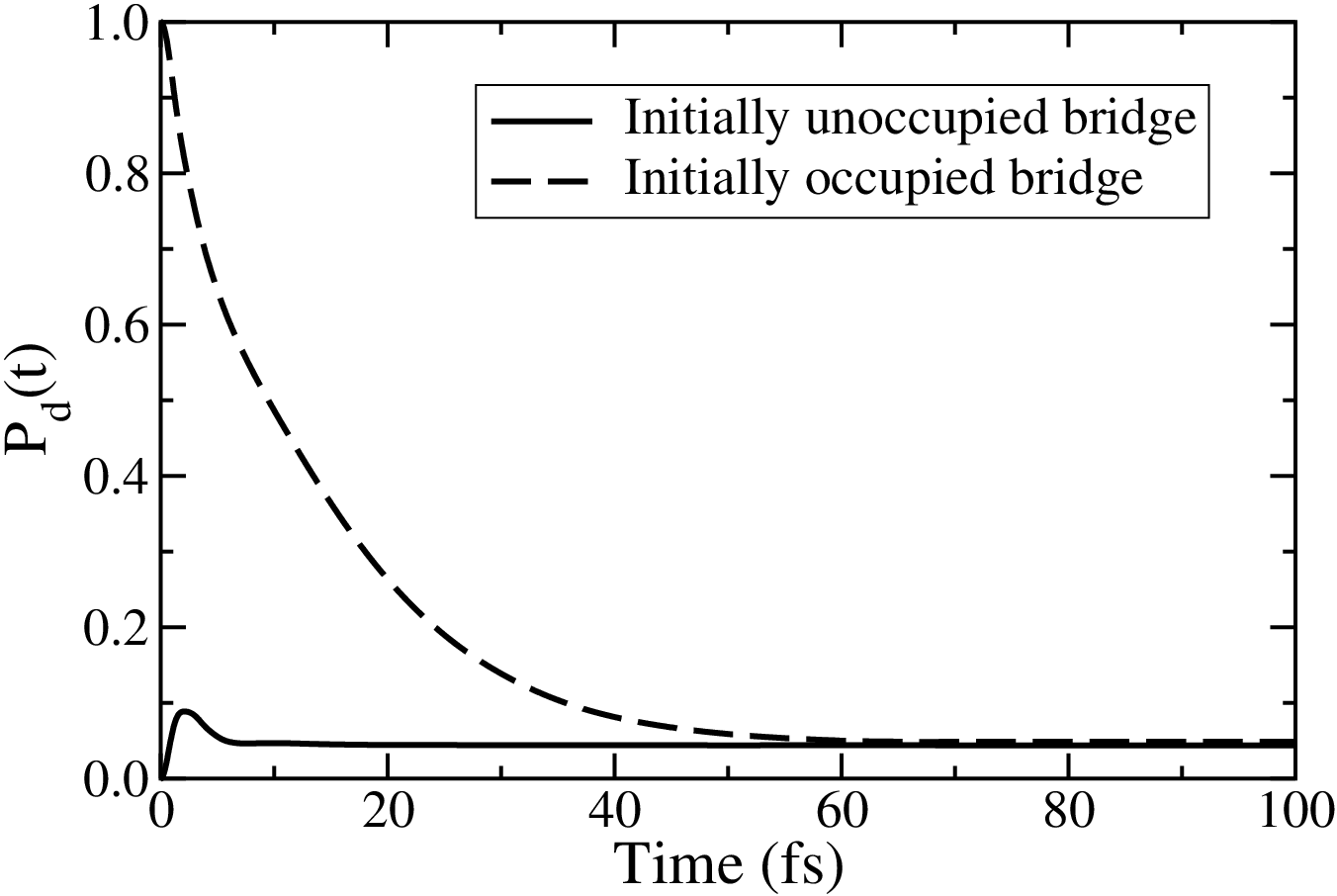}

\caption{Dependence of the time-dependent electric current (a) and the
  population of the bridge state (b) on the initial state.
The electronic parameters are  $\alpha_e = 0.2$eV, $\beta_e = 1$eV, $E_d - E_f = 0.5$eV,
and $V = 0.1$V. The vibrational parameters are $\lambda = 2000 {\rm cm}^{-1}$ and 
$\omega_c = 500{\rm cm}^{-1}$.}
\label{fig5}
\end{figure}

For the model parameters considered in this paper, our calculations show that the stationary state is
independent on the initial condition. As an  example, 
Figure~\ref{fig5}(a)  shows that  the two initial states discussed above indeed give the same stationary current 
for the electronic parameters  $\alpha_e = 0.2$ eV, $\beta_e = 1$ eV, $E_d -
E_f = 0.5$ eV,
$V = 0.1$ V, and the vibrational parameters  $\lambda = 2000$ ${\rm cm}^{-1}$ and 
$\omega_c = 500$ ${\rm cm}^{-1}$, despite the fact that their initial transient
characteristics are quite different.
As illustrated in Figure ~\ref{fig5}(b), this is due to the fact that although the initial
bridge state populations $P_d(t)$, defined by Eq.\ (\ref{population}), are quite different, they attain the same stationary value within 
a relatively short time scale.  
For other parameters, however,  the stationary state may depend on the initial
condition. The
investigation of the corresponding phenomenon of
bistability\cite{Galperin05,Alexandrov07,Galperin08,Ryndyk08,DAmico08,Riwar09} 
will be the subject of future work.\cite{note1}

We note that different sets of initial conditions also affect the
time scale at which the current  $I(t)$ reaches its stationary value, 
as is evident from Figure~\ref{fig5}.  In our
simulations we typically choose initial conditions that are close to the final steady state, e.g., an
unoccupied initial bridge state with an unshifted bath of oscillators for the transport calculations
of Fig.~\ref{fig4}(a) and an occupied bridge state with a bath of oscillators in equilibrium with the 
occupied bridge state for calculations of Fig.~\ref{fig4}(b). 

\begin{figure}[!ht]

\includegraphics[clip,width=0.45\textwidth]{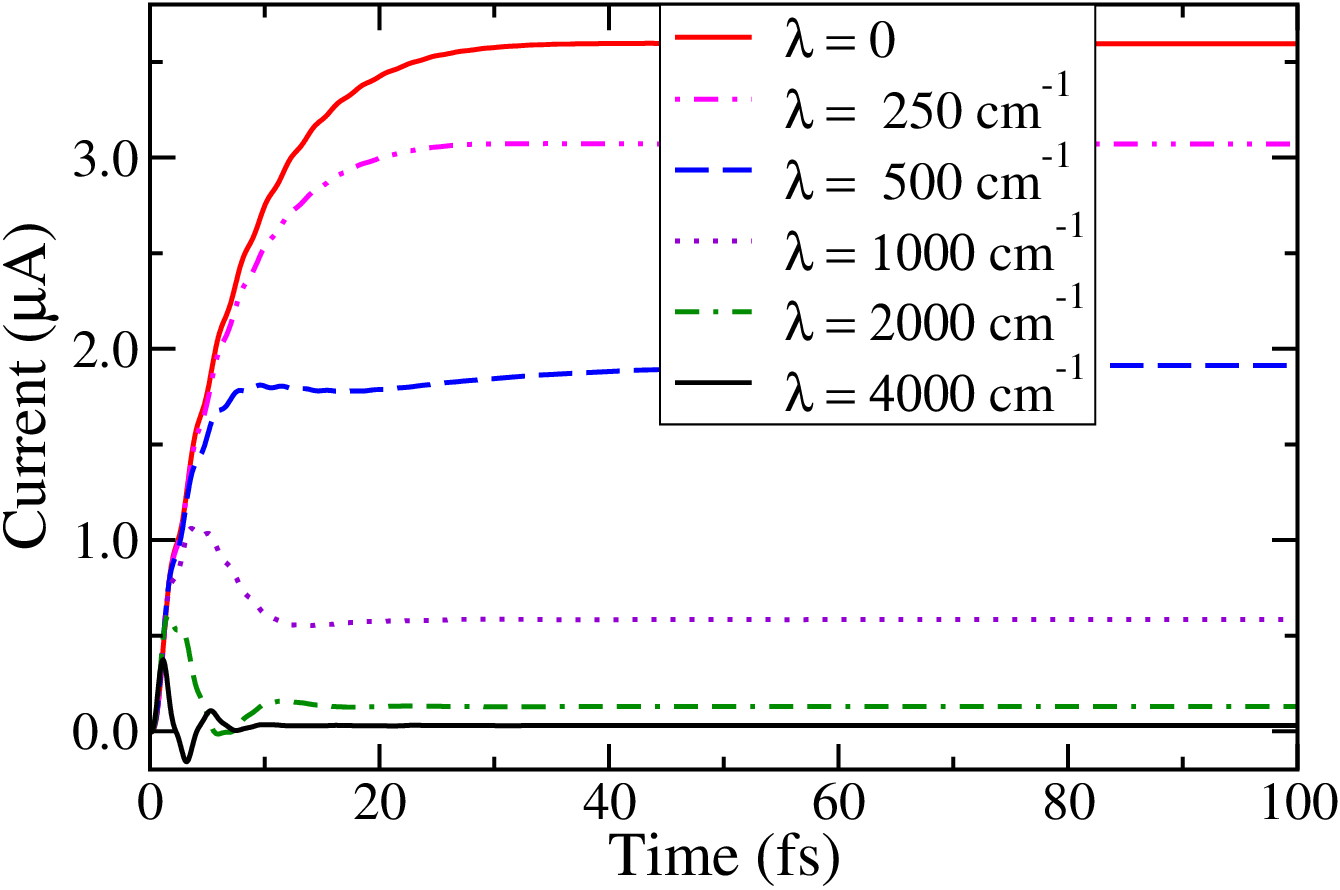}

\caption{Time-dependent current $I(t)$ in the phonon-blockade regime 
for different coupling strengths to the vibrational
  bath as specified by the reorganization energy $\lambda$.
The electronic parameters are $\alpha_e = 0.2$eV, $\beta_e = 1$eV, $E_d - E_f = 0$,
and $V = 0.1$V. The characteristic frequency of the vibrational bath is $\omega_c = 500{\rm cm}^{-1}$.}
\label{fig6}
\end{figure}

\begin{figure}[!ht]

\includegraphics[clip,width=0.45\textwidth]{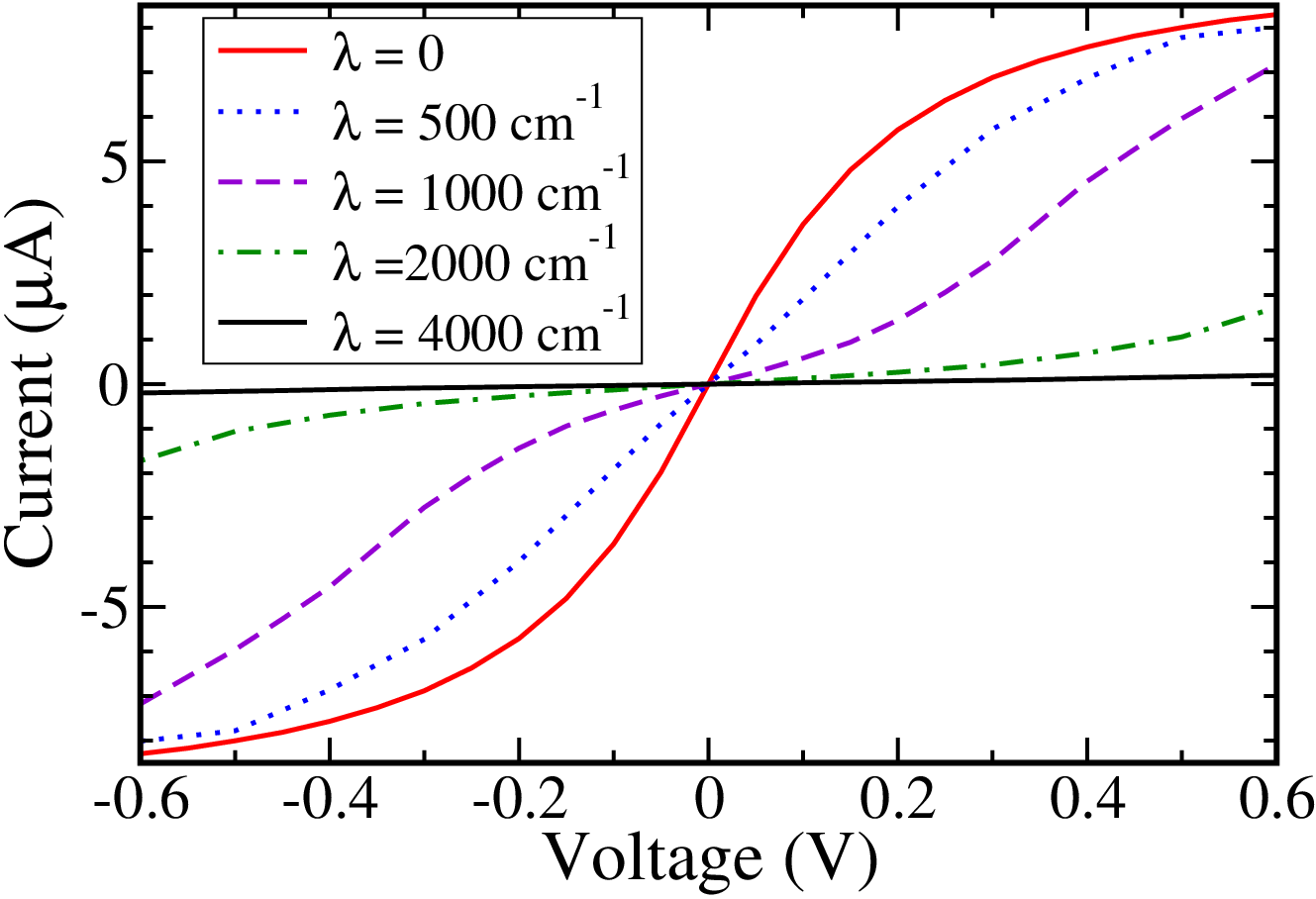}

\caption{Dependence of the current-voltage characteristics in the
  phonon-blockade regime on the electronic-vibrational
coupling strength. The electronic parameters are the same as in Fig.~\ref{fig6}, with
$\omega_c = 500{\rm cm}^{-1}$.}
\label{fig7}
\end{figure}

We finally consider a model where the energy of the bridge state is located
at the Fermi energy of the leads.
This parameter regime is particularly interesting, because already for small
bias voltage the transport mechanism corresponds to resonant tunneling and involves
mixed electron/hole transport. 
For a purely electronic model, the Landauer formula predicts the maximum current when $E_d = E_f$.  
Including the couplings to the vibrational modes, however, may have a significant impact on the 
electric current. This is illustrated in Figure~\ref{fig6} for different
electronic-vibrational coupling strengths $\lambda$. 
It is seen that for short time the current $I(t)$ obtained for finite
$\lambda$ follows the  current for the purely electronic model ($\lambda =
0$). However, after a short transient time the coupling to the vibrations
becomes effective and results in a suppression of the current. 
In particular for larger vibrational coupling,  $\lambda = 2000-4000$ ${\rm cm}^{-1}$, the
effect is very pronounced and the stationary current is essentially blocked
over a significant range of bias voltages, as is demonstrated by
the current-voltage characteristics in Figure~\ref{fig7}. 

The underlying  mechanism can be qualitatively rationalized by considering the
energy level of the bridge state. For any finite bias voltage, the bare energy of the
bridge state ($E_d - E_f = 0$) is located between  the chemical potential of
the leads and thus, within a purely electronic model, current can
flow. The coupling to the vibrations results in a polaron shift of the energy of the bridge state. For
electronic-vibrational coupling strengths  $\lambda > |V|/2$ the
polaron-shifted energy of the
bridge state is below the chemical potentials of both leads and thus
current is blocked. This effect, referred to as phonon blockade of the current,
has been observed e.g.\ in
quantum dots.\cite{Weig04}

\begin{figure}[!ht]

\includegraphics[clip,width=0.45\textwidth]{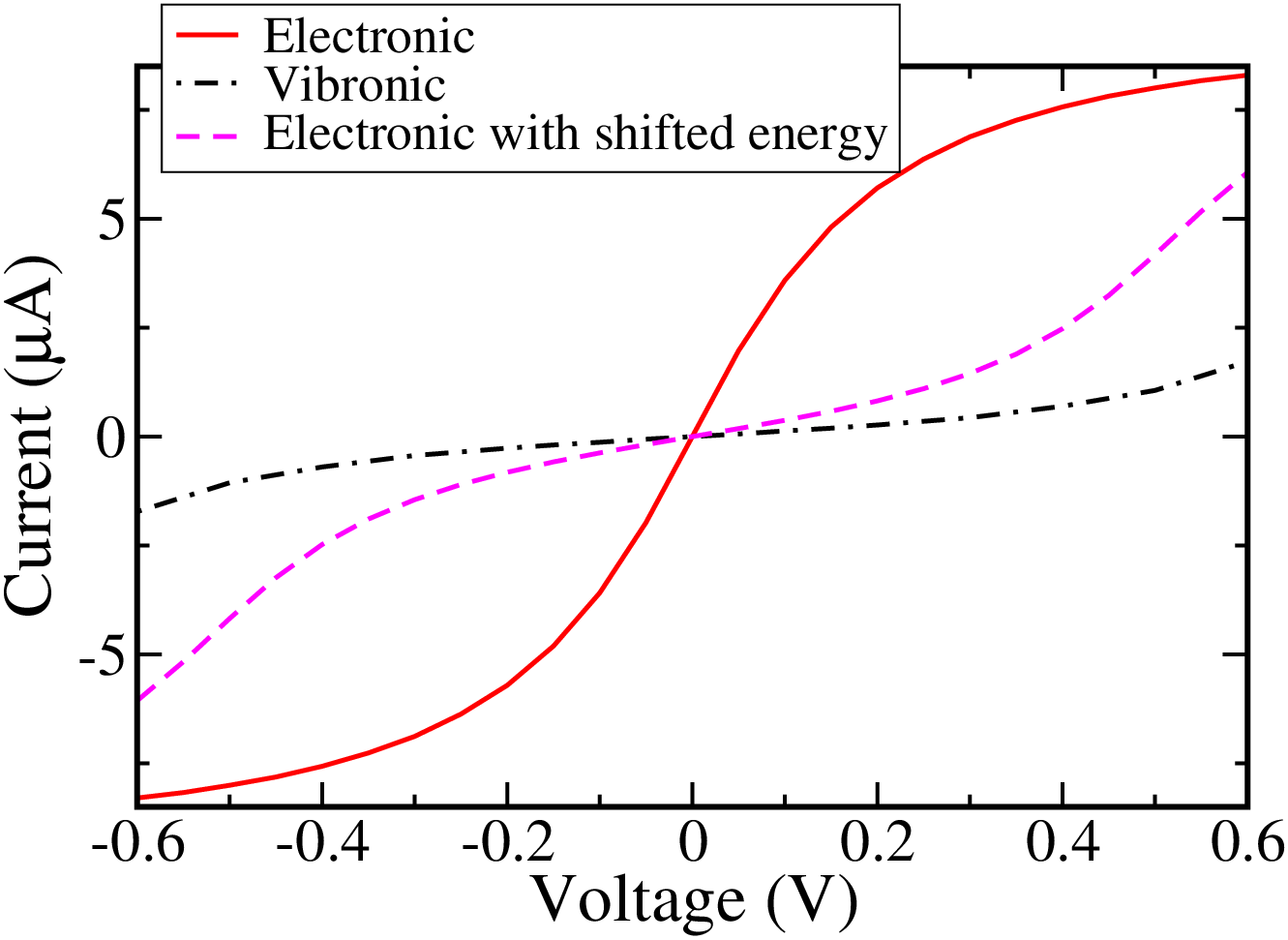}

\caption{Current-voltage characteristics for vibrationally coupled electron
  transport in the phonon-blockade regime. The electronic parameters 
are the same as in Fig.~\ref{fig6}.  The vibrational parameters are $\lambda = 2000 {\rm cm}^{-1}$
and $\omega_c = 500{\rm cm}^{-1}$. Shown are results of a purely electronic
model  employing the bare (full line) and the polaron-shifted (dashed line) energy of the discrete
state, respectively, as well as results of a full vibrationally-coupled many-body ML-MCTDH-SQR
calculation (dashed-dotted line).}
\label{fig8}
\end{figure}

Although the interpretation  of the phonon blockade in terms of the energetics
of the bridge state is appealing,
it should be emphasized that the mechanism of phonon blockade involves the
formation of a many-body polaron-type state that is significantly more complex
than this  purely electronic picture and cannot be
fully described by just considering the static shift of the energy of the bridge state. 
The bare energy and the polaron-shifted energy of the
bridge state are only two special points on the multidimensional potential
energy surface of the charged state given by 
$V_d({\bf Q}) = E_d +\frac{1}{2} \sum_j\omega_j^2 Q_j^2 + \sum_j c_j Q_j$.
For values  $\lambda > |V|/2$ the potential energy surface of
the discrete state crosses the  chemical potential of the leads 
as a function of the nuclear coordinates $Q_j$. In this parameter regime, 
an accurate description of the vibrational dynamics
and its coupling to the electronic degrees of freedom is required to
obtain a quantitative description of the many-body polaron state and its transport characteristics. 
This is demonstrated in Figure~\ref{fig8}, which compares the electric
current obtained with a full vibrationally-coupled  many-body ML-MCTDH-SQR calculation to that
obtained with a purely electronic model for a polaron-shifted energy of the
bridge state, i.e.\ $E_d \rightarrow E_d - \lambda$.
The comparison shows that the effect
cannot be described properly with a purely electronic model but requires
methods that incorporate the dynamical effect of the vibrations on the transport.

\begin{figure}[!ht]
\begin{flushleft}
(a)
\end{flushleft}
\includegraphics[clip,width=0.4\textwidth]{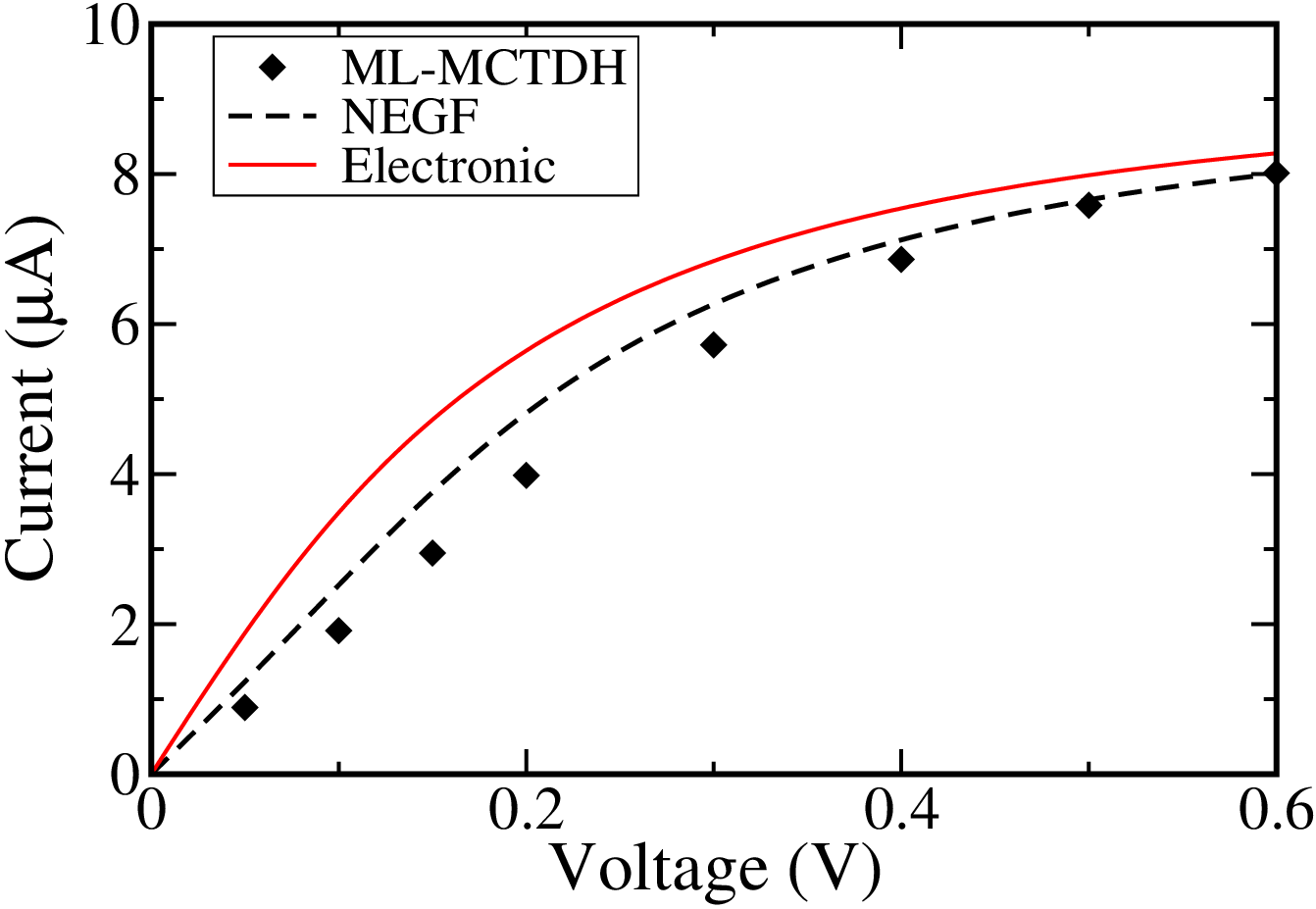}

\begin{flushleft}
(b)
\end{flushleft}
\includegraphics[clip,width=0.4\textwidth]{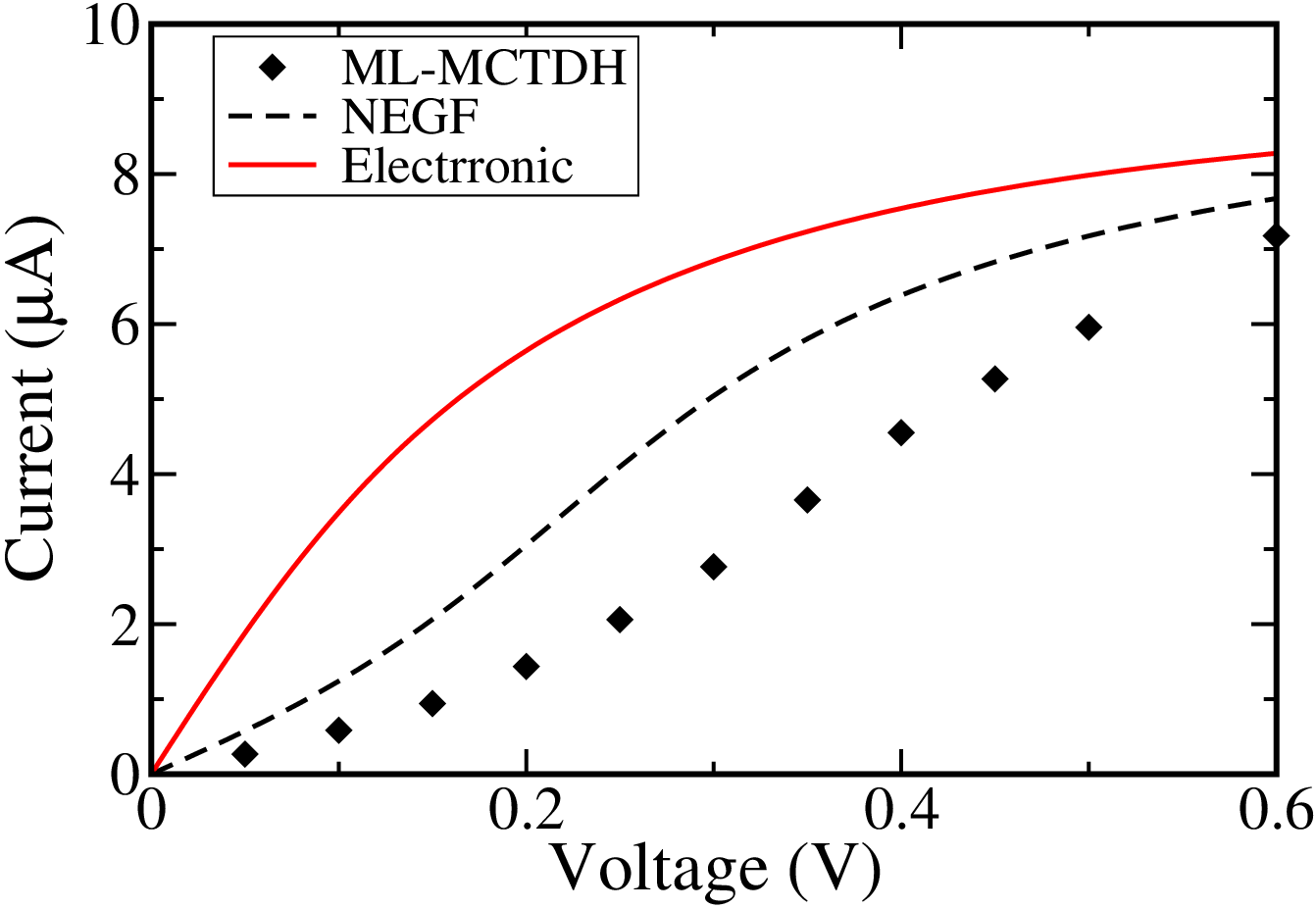}

\begin{flushleft}
(c)
\end{flushleft}
\includegraphics[clip,width=0.4\textwidth]{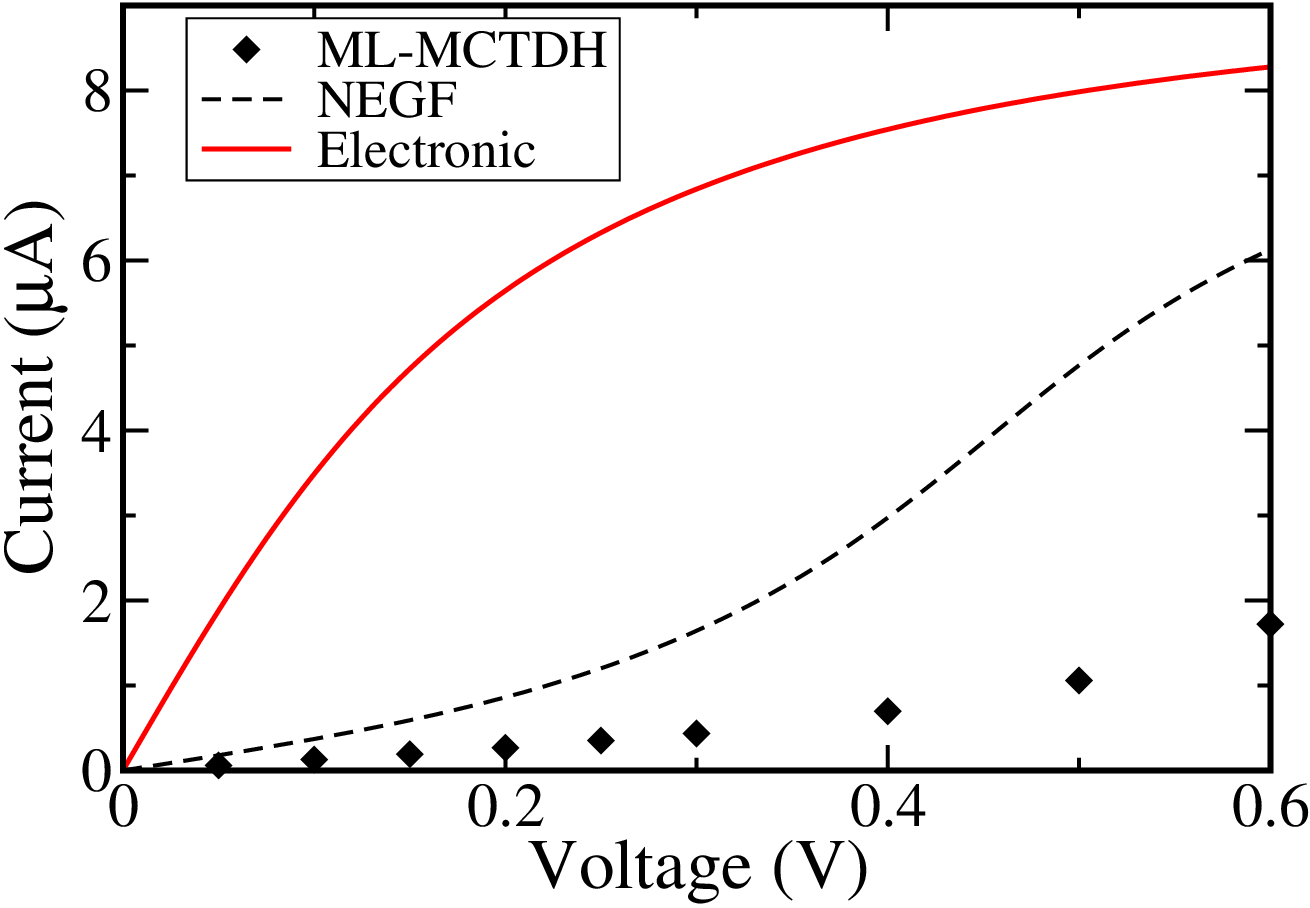}

\caption{Comparison of results obtain with NEGF theory (dashed lines) and the ML-MCTDH-SQR
  method (diamonds) for vibronic coupling parameters   $\lambda = 500 {\rm
    cm}^{-1}$ (a), $\lambda = 1000 {\rm cm}^{-1}$ (b), and  $\lambda = 2000
  {\rm cm}^{-1}$ (c). All other parameters are the same as in Fig.~\ref{fig6}. 
In addition, results for a purely electronic model are
  shown (full lines).}
\label{fig9}
\end{figure}

As discussed in the introduction, a variety of approximate methods have been
developed and employed to describe vibrationally coupled electron transport in
molecular junctions, including scattering
theory,\cite{Bonca95,Ness01,Cizek04,Cizek05,Toroker07,Benesch08,Zimbovskaya09,Seidemann10} 
nonequilibrium Green's function (NEGF)
approaches,\cite{Flensberg03,Mitra04,Galperin06,Ryndyk06,Frederiksen07,Tahir08,Haertle08,Stafford09,Haertle09} 
and master equation
methods.\cite{May02,Mitra04,Lehmann04,Pedersen05,Harbola06,Zazunov06,Siddiqui07,Timm08,May08,May08b,Leijnse09,Esposito09,Haertle11} 
However, only very few of them are applicable to
the present model. This is because the model involves vibronic
coupling  to a relatively large number of vibrational modes
in nonequilibrium. Master equation methods are limited to a small number
of vibrational degrees of freedom that are treated in full nonequilibrium, while 
many scattering theory approaches neglect vibrational nonequilibrium
effects. NEGF theory is, in principle, applicable to describe coupling to a larger
number of harmonic  vibrational modes in nonequilibrium. However, if implemented 
within the self-consistent Born approximation  it is limited to small electron-vibrational coupling  and has
so far mostly been applied in the nonresonant tunneling regime.\cite{Frederiksen07}
To treat vibrationally coupled electron transport in the resonant regime,
another NEGF method has been 
proposed by Galperin et al.\cite{Galperin06} and extended by
H\"artle et al.\cite{Haertle08,Haertle09,Volkovich11}  Being based on the
polaron transformation, this NEGF method is in principle able to treat
moderate vibronic coupling strengths. We have applied this method to the
present model employing 10 vibrational modes to model the vibrational
distribution described by the spectral density, Eq.~(\ref{ohmic}).\cite{note1}
Fig.~\ref{fig9} shows a comparison of results
of the NEGF method for the stationary current with those obtained from ML-MCTDH for different vibronic
coupling strength. Overall, the results indicate that the NEGF method is
capable of describing the suppression
of the current due to phonon blockade. For small vibronic
coupling the NEGF results are in almost quantitative agreement with the numerically exact
results. However, for larger vibronic coupling NEGF theory underestimates the effect of phonon
blockade. This is presumably due to the fact that the present model does not
exhibit a strict time-scale separation between the electronic and nuclear
degrees of freedom, which is a prerequisite for the NEGF method. These results
demonstrate that with approximate methods
the simulation of transport properties of the present model, which is in the
nonperturbative regime,   is challenging.  A more
general validation of the NEGF method in a broader regime requires extensive
studies by both the ML-MCTDH-SQR and the NEGF methods and will be the subject of future work.

\begin{figure}[!ht]

\includegraphics[clip,width=0.45\textwidth]{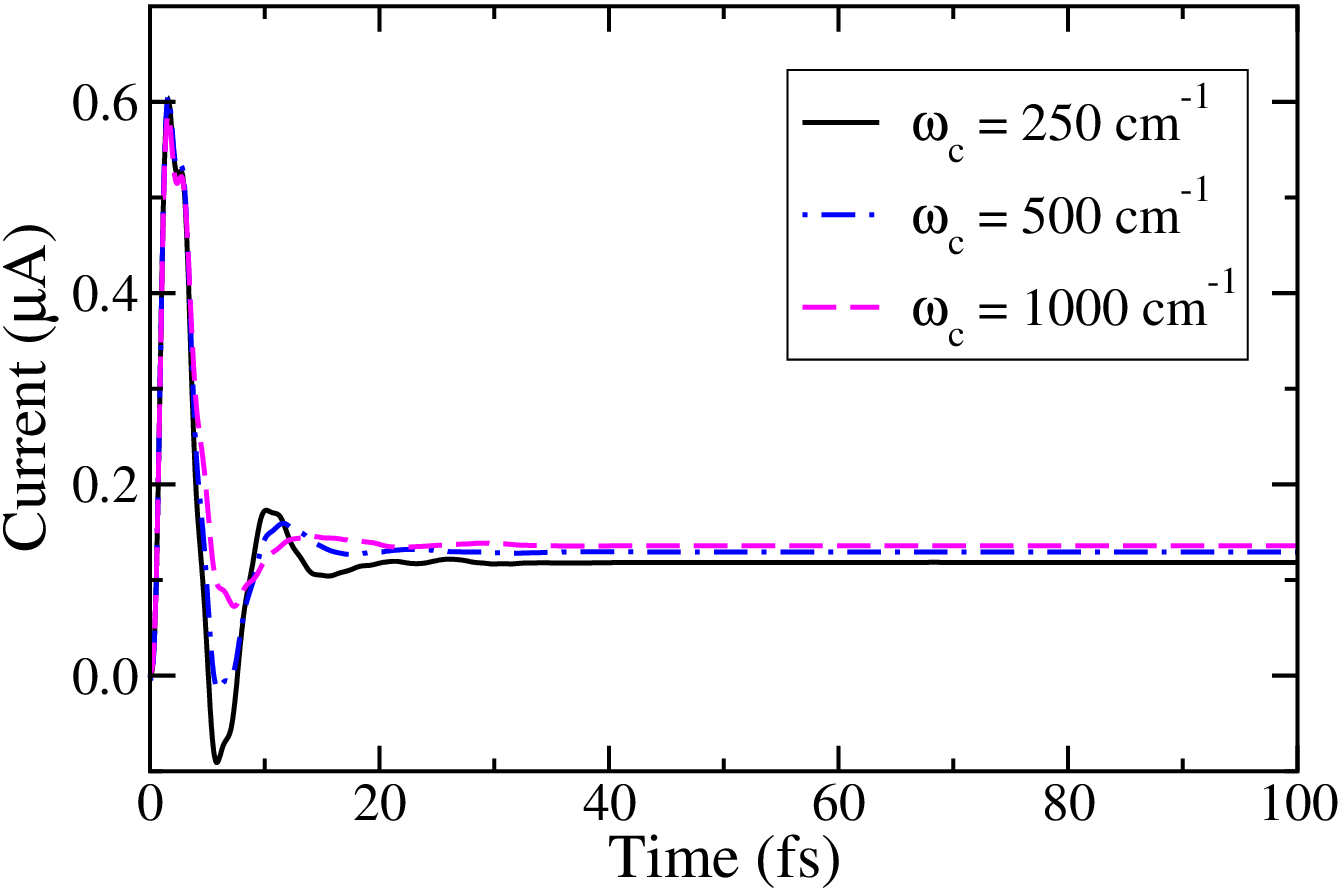}

\caption{Dependence of the time-dependent current $I(t)$ in the phonon-blockade
  regime on the characteristic frequency of the vibrational bath.
The electronic parameters are $\alpha_e = 0.2$eV, $\beta_e = 1$eV, $E_d - E_f = 0$,
and $V = 0.1$V. The reorganization energy is $\lambda = 2000 {\rm cm}^{-1}$.}
\label{fig10}
\end{figure}

While the effect of  phonon blockade of the stationary current is to be expected for
energetic reasons, the treatment with the ML-MCTDH method also allows 
a detailed the study of the time-dependent formation of the
underlying many-body polaron state. For example, Figure~\ref{fig10} shows that the
transient dynamics depends significantly on the
characteristic frequency of the vibrational bath, $\omega_c$, whereas the
value of the stationary current is relatively insensitive to $\omega_c$.  
This is due to the fact that the frequency $\omega_c$ determines the timescale
on which the system moves on the potential energy surface from the initially
prepared state, corresponding to the bare energy of the bridge states, to the
relaxed state below the chemical potential of the leads.

It is worthwhile to emphasize that the mechanism of the phonon blockade analyzed here is different
from that of
the previously discussed Franck-Condon blockade,\cite{Koch05,Koch06,Leturcq09} which also leads to
a suppression of the current due to strong electronic-vibrational
coupling. While the former can be removed by a gate potential that shifts the
energy of the bridge state into the conductance window, 
i.e.\ between the chemical potential of the electrodes,
the latter is rather insensitive to a gate potential.

We finally discuss some technical details of the numerical calculations employing
the ML-MCTDH-SQR method. For the parameter regimes investigated in this paper, 
the stationary current is usually reached 
at approximately 20-60 fs.  To ensure convergence most calculations were carried out to 100 fs. 
Within this time scale of simulation, 64-128 discrete states are used to represent each lead's 
electronic continuum, resulting in a total number of 64-128 electrons in the ML-MCTDH-SQR 
numerical treatment. The nuclear bath is represented by 100-400 modes.  The converged number 
of basis functions for these vibrational modes ranges from a few to a few hundred.  The
calculation was performed with a four-layer ML-MCTDH-SQR theory, with one top-layer SP group for
the vibrational modes and three top-layer SP groups for the electronic part.
These SP groups are then recursively expanded via a binary tree (i.e., two lower-layer SP groups 
in each preceding upper-layer SP).  The final converged results (to within 10\% relative error) 
require 40-60 SPFs for each top-layer electronic SP group, 20-40 SPFs for all lower-layer 
electronic SPs, and 10 SPFs for all nuclear SP groups.  This results in a total of $\sim 10^6$ 
equations to solve.  Each simulation took between 20 hours and several days of CPU time on a 
typical personal computer. Calculations for a finite temperature bath requires ensemble average 
over a few hundred initial wave functions, and were performed on a Cray XT4 parallel computer.  

\begin{figure}[!ht]
\begin{flushleft}
(a)
\end{flushleft}
\includegraphics[clip,width=0.45\textwidth]{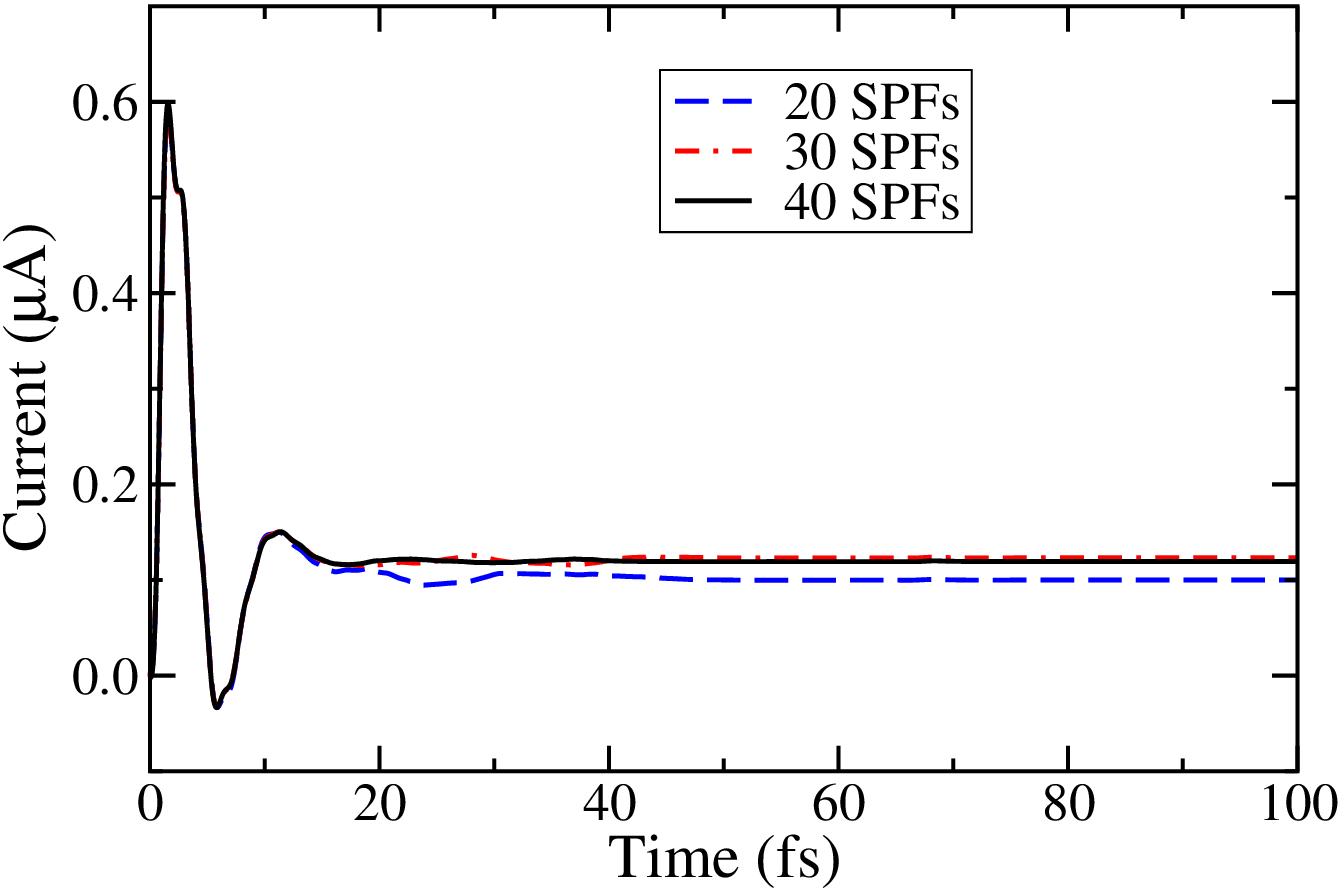}

\begin{flushleft}
(b)
\end{flushleft}
\includegraphics[clip,width=0.45\textwidth]{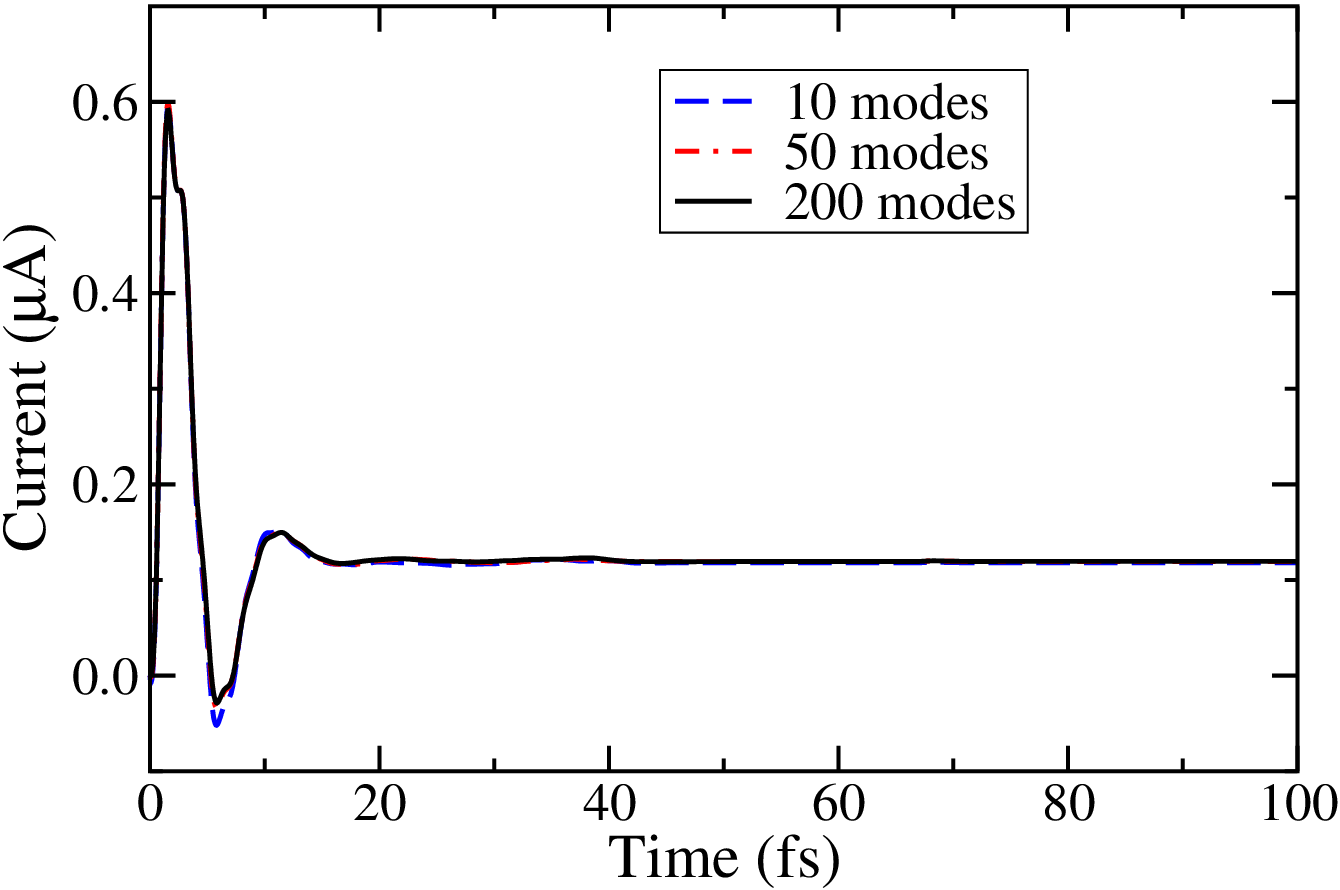}

\begin{flushleft}
(c)
\end{flushleft}
\includegraphics[clip,width=0.45\textwidth]{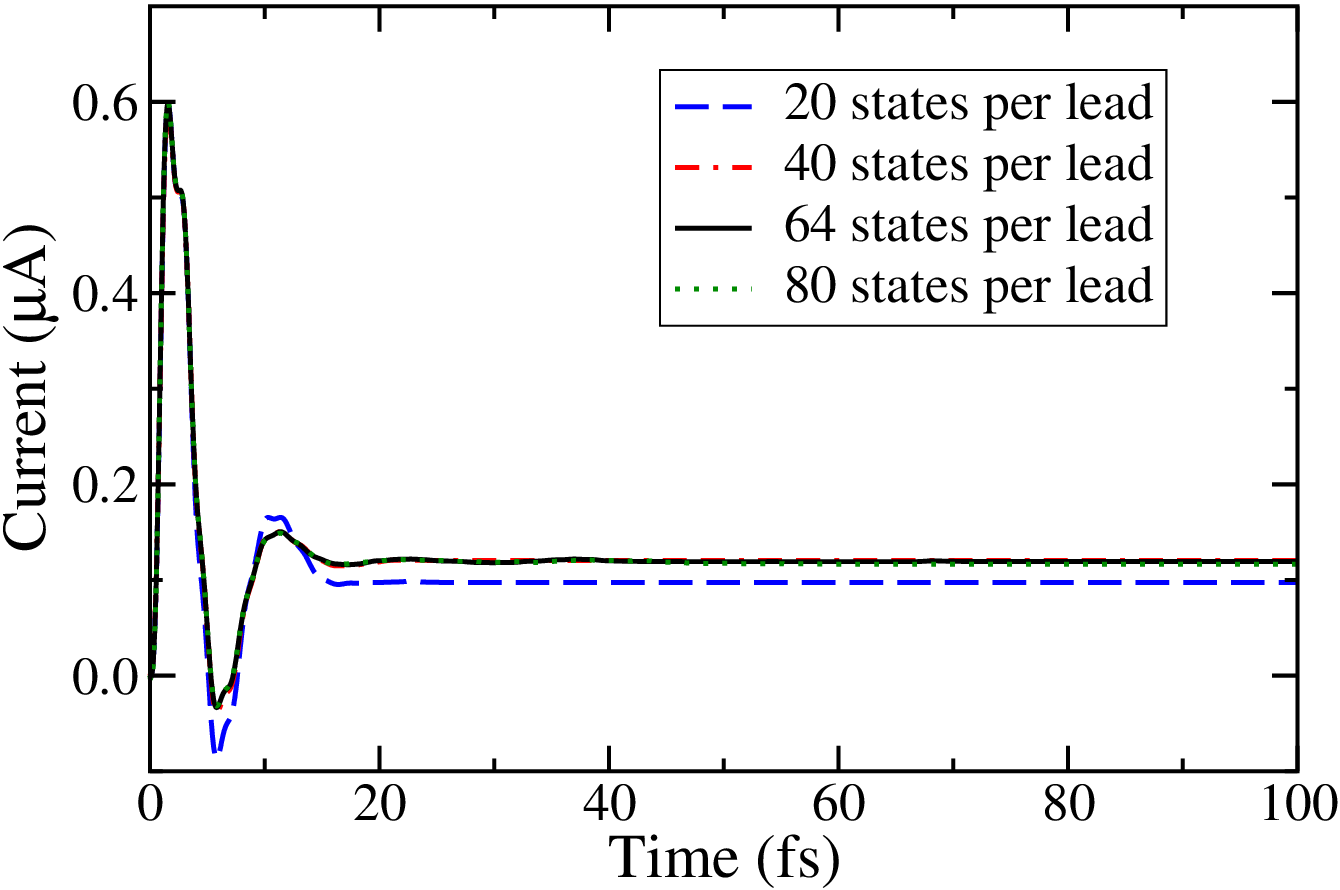}

\caption{Time-dependent current $I(t)$ for the parameter set of
$\alpha_e = 0.2$eV, $\beta_e = 1$eV, $E_d - E_f = 0.5$eV, and $V = 0.1$V. 
Convergence is shown with respect to: (a) the number of SPFs for the electronic SPs;
(b) the number of bath modes; (c) the number of electronic states for each lead.
Other variational parameters are described in the text.  }
\label{fig11}
\end{figure}

The convergence of the ML-MCTDH-SQR simulation is illustrated in Figure~\ref{fig11},
where the physical parameters are the same as those in Fig.~\ref{fig1} except for a
different voltage of 0.1V.  The calculations are performed with a four-layer ML-MCTDH-SQR
scheme. For simplicity the convergence is shown for the following three
different categories.  In Figure~\ref{fig11}(a) there are 64 electronic states for each 
lead and 50 modes for the nuclear bath.  The number of the single particle functions (SPFs) 
is 10 for each nuclear SP group (of each layer), and the number of the SPFs for each electronic SP 
group (of each layer) is set the same and varied.  In Figure~\ref{fig11}(b) there are 64 electronic 
states for each lead, with 40 SPFs for each electronic SP group.  The number of bath modes is varied, 
with 10 SPFs for each SP group of each layer. In Figure~\ref{fig11}(c) there are 50 bath modes, 
with 10 SPFs for each SP group.  The number of electronic states for each lead is varied, with 40 
SPFs for each SP group (of each layer).  

First, we consider the convergence with respect to the number of SPFs. 
Fig.~\ref{fig11}(a) shows that the electronic
(fermionic) degrees of freedom require a relatively large number of SPFs to achieve convergence.  
The steady state current obtained with 20 SPFs differs approximately 20\% from the converged value, 
although the short time transient dynamics agrees well with other results obtained with
a larger number of SPFs.  In this case convergence is reached when the number of SPFs 
exceeds 30 for the electronic degrees of freedom.  For the nuclear degrees of freedom, 
tests have shown that results obtained with 6 or 8 SPFs are nearly identical to that 
obtained with 10 SPFs.  Thus, although we used 10 SPFs in all the calculations, we believe
a smaller number could be equally satisfactory.

Next, we check convergence with respect to the number of bath modes.  Figure~\ref{fig11}(b) shows
that only small differences are found when the number of modes changes from 10 to 200.  If one
is only interested in the steady state current and not the finer details of the transient
dynamics of $I(t)$, then a bath of 10 modes is sufficient as is used in our NEGF calculations.  On
the other hand, to represent the electronic continuum within the timescale of simulation,
a sufficient number of electronic states are required.  As shown in Fig.~\ref{fig11}(c), 20
states per lead is inadequate for both the transient $I(t)$ or the steady state current. For
the case of Fig.~\ref{fig11}  convergence is achieved when the number of states per lead is greater
than 40.

\section{Concluding Remarks}\label{conclusions}

In this paper we have employed the ML-MCTDH-SQR method to simulate vibrationally
coupled electron transport through single-molecule junctions. The ML-MCTDH-SQR
method allows an accurate, in principle numerically exact treatment of this many-body
quantum transport problem. 
The results obtained for a generic model demonstrate the importance
of electronic-vibrational coupling, which has a significant influence on the
transport properties.
For situations where the energy of the bridge state is located close to the
Fermi energy, the simulations show the time-dependent formation of a polaron
state that results in a pronounced suppression of the current corresponding to
the phenomenon of phonon blockade. We have shown that this phenomenon cannot
be explained solely by the polaron shift
of the energy but requires methods that incorporate the dynamical effect of
the vibrations on the transport.

While some of these results have been discussed previously based on
approximate methods, the present methodology does not involve any systematic
approximations and provides accurate benchmark results. It can thus also been
used to test the validity of more approximate methods. As an example, we have
have discussed the validity of a NEGF method in the parameter regime of phonon blockade, 
demonstrating that accurate methods such as ML-MCTDH are necessary to study transport in
the strong coupling regime.
A more detailed study of the validity of approximate methods as well as
the application of the methodology to investigate signatures of vibronic effects in 
experimental transport spectra, such as, e.g., inelastic electron tunneling spectroscopy, 
will be the subject of future work.

It is also emphasized that
the time-dependent treatment employed in the ML-MCTDH-SQR method
provides not only information on the steady state but also on the transient 
dynamics and can thus also been used to
study the influence of time-dependent electrical fields, such as, e.g., ac gate fields
or optical pulses on the transport process.

In the present study we have focused on the effect of electronic-vibrational
coupling on transport in molecular junctions. Another important mechanism is
electron-electron interaction. The extension of the ML-MCTDH-SQR to include
explicit electron-electron interaction is currently under way. 
This may open the perspective to a comprehensive many-body treatment of
nonequilibrium charge transport at the nanoscale.

\section*{Acknowledgments}
This work has been supported by the National Science Foundation
CHE-1012479 (HW) and the Deutsche Forschungsgemeinschaft (DFG) through  a research grant and
the Cluster of Excellence Munich Center of Advanced Photonics (MT), and used resources
of the Leibniz Rechenzentrum M\"unchen, the Rechenzentrum Erlangen (RRZE), and the 
National Energy Research Scientific Computing Center, which is supported by the
Office of Science of the U.S.  Department of Energy under Contract No. DE-AC02-05CH11231.

\pagebreak

%
%


\clearpage

\clearpage
~
\vspace{3cm}

\clearpage
~
\vspace{3cm}

\clearpage

\clearpage

\clearpage

\vspace*{3cm}

\clearpage
~
\vspace{3cm}

\clearpage
~
\vspace{3cm}

\clearpage

\clearpage

\vspace*{3cm}

\clearpage

\end{document}